\documentstyle[12pt,axodraw,epsfig]{article}
\begin{document}
\baselineskip 0.6cm

\def\simgt{\mathrel{\lower2.5pt\vbox{\lineskip=0pt\baselineskip=0pt
           \hbox{$>$}\hbox{$\sim$}}}}
\def\simlt{\mathrel{\lower2.5pt\vbox{\lineskip=0pt\baselineskip=0pt
           \hbox{$<$}\hbox{$\sim$}}}}
\def\L{\kappa}

\begin{titlepage}

\begin{flushright}
LBNL-54212 \\
UCB-PTH-03/36
\end{flushright}

\vskip 2.0cm

\begin{center}

{\Large \bf 
Holographic Theories of Electroweak Symmetry Breaking 
without a Higgs Boson
}

\vskip 1.0cm

{\large
Gustavo Burdman$^a$ and Yasunori Nomura$^{a,b}$
}

\vskip 0.4cm

$^a$ {\it Theoretical Physics Group, Lawrence Berkeley National Laboratory,
                Berkeley, CA 94720}\\
$^b$ {\it Department of Physics, University of California,
                Berkeley, CA 94720}

\vskip 1.2cm

\abstract{
Recently, realistic theories of electroweak symmetry breaking have been 
constructed in which the electroweak symmetry is broken by boundary 
conditions imposed at a boundary of higher dimensional spacetime. 
These theories have equivalent 4D dual descriptions, in which the 
electroweak symmetry is dynamically broken by non-trivial infrared 
dynamics of some gauge interaction, whose gauge coupling $\tilde{g}$ 
and size $N$ satisfy $\tilde{g}^2 N \simgt 16\pi^2$.  Such theories 
allow one to calculate electroweak radiative corrections, including the 
oblique parameters $S$, $T$ and $U$, as long as $\tilde{g}^2N/16\pi^2$ 
and $N$ are sufficiently larger than unity.  We study how the duality 
between the 4D and 5D theories manifests itself in the computation of 
various physical quantities.  In particular, we calculate the electroweak 
oblique parameters in a warped 5D theory where the electroweak symmetry 
is broken by boundary conditions at the infrared brane.  We show that 
the value of $S$ obtained in the minimal theory exceeds the experimental 
bound if the theory is in a weakly coupled regime.  This requires either 
an extension of the minimal model or departure from weak coupling. 
A particularly interesting scenario is obtained if the gauge couplings 
in the 5D theory take the largest possible values -- the value suggested 
by naive dimensional analysis.  We argue that such a theory can provide 
a potentially consistent picture for dynamical electroweak symmetry 
breaking: corrections to the electroweak observables are sufficiently 
small while realistic fermion masses are obtained without conflicting 
with bounds from flavor violation.  The theory contains only the standard 
model quarks, leptons and gauge bosons below $\simeq 2~{\rm TeV}$, except 
for a possible light scalar associated with the radius of the extra 
dimension.  At $\simeq 2~{\rm TeV}$ increasingly broad string resonances 
appear.  An analysis of top-quark phenomenology and flavor violation 
is also presented, which is applicable to both the weakly-coupled 
and strongly-coupled cases.
}

\end{center}
\end{titlepage}

\section{Introduction}
\label{sec:intro}

One of the greatest mysteries in particle physics is the origin of 
electroweak symmetry breaking.  In the standard model the electroweak 
symmetry is broken by a vacuum expectation value (VEV) of a Higgs field, 
which is driven by a non-trivial potential introduced to break the 
symmetry.  However, once the theory is extrapolated to higher energies 
in a perturbative way, one finds that the Higgs mass squared parameter 
receives large radiative corrections of the order of the cutoff scale, 
destabilizing the electroweak scale.  Therefore, it is quite natural 
to suspect that some non-trivial strong dynamics is responsible for 
electroweak symmetry breaking in a direct or indirect way.  Such a 
consideration leads to theories where the electroweak symmetry is broken 
by a condensation caused by strong gauge dynamics~\cite{Weinberg:gm} 
or theories where the Higgs boson arises as a composite state of some 
strong interaction~\cite{Kaplan:1983fs}.  In these theories the strength 
of the relevant gauge interaction is weaker at higher energies, and
becomes non-perturbative only at lower energies by the renormalization 
group evolution.  This triggers electroweak symmetry breaking at 
exponentially lower energies compared with the cutoff scale, thus 
evading the problem of the stability. 

In this paper we explore alternative possibilities for ``dynamical'' 
theories of electroweak symmetry breaking.  Suppose the gauge interaction, 
which is responsible for electroweak symmetry breaking and non-perturbative 
at the electroweak scale, stays very strong at higher energies -- 
stronger than that above which the conventional perturbation theory 
breaks down.  Apparently, this does not provide any viable description 
of physics at energies higher than the electroweak scale. However, the 
presence of dualities between the 4D gauge theories and higher dimensional 
gravitational theories suggests that such a theory is described, 
in fact, by a higher dimensional theory where the electroweak symmetry 
is broken by the presence of a spacetime boundary.  This relation becomes 
particularly concrete when the theory on the gravitational side is on 
an Anti~de-Sitter (AdS) background~\cite{Maldacena:1997re}, and it has been 
used to build models of a composite Higgs boson~\cite{Contino:2003ve} and 
dynamical electroweak symmetry breaking~\cite{Csaki:2003zu,Nomura:2003du}. 
However, the background geometry of the gravitational theory may not 
necessarily be AdS, as in the models considered in~\cite{Csaki:2003dt,%
Barbieri:2003pr}.

In this paper we consider theories of the kind described above, in which 
the holographic description of the theory relates a higher dimensional 
theory to some 4D ``gauge theory''.  In particular, we study theories 
where the electroweak symmetry is broken ``dynamically'' without the 
presence of the physical Higgs boson --- in the higher dimensional picture 
this corresponds to the theories where the electroweak symmetry is broken 
by boundary conditions imposed at a boundary of the spacetime.  We mainly 
consider theories formulated in the AdS space, in which the electroweak 
symmetry is broken by boundary conditions imposed at the infrared (IR) 
brane~\cite{Csaki:2003zu,Nomura:2003du}, but some of our analysis applies 
to more general theories such as the ones in flat space~\cite{Csaki:2003dt,%
Barbieri:2003pr}.  In the actual analysis we adopt the specific theory 
constructed in~\cite{Nomura:2003du}, which reproduces many successful 
features of the standard model including fermion mass generation and 
suppression of flavor changing neutral currents (FCNCs).  This theory also 
allows us to control the scale of new physics, which corresponds in the 4D 
picture to the size (the number of ``colors'') of the gauge interaction, 
and thus represents a class of generic theories in 5D AdS space. 
We study electroweak radiative corrections and find that the constraints 
from precision electroweak measurements prefer gauge groups of smaller 
size, unless some additional contribution to the electroweak oblique 
parameters is introduced.  This situation is similar to that in 
technicolor theories~\cite{Peskin:1990zt}.  We elucidate how such 
a similarity arises in general theories with the electroweak symmetry 
broken by boundary conditions.

Although regarding electroweak corrections the situation in our theory 
is similar to that in technicolor, other aspects can be quite different. 
In particular, we expect that the theory does not have problems in 
general to obtain realistic fermion masses, correct vacuum alignment, 
and suppression of flavor violation.  This implies that even the minimal 
theory may have a viable parameter region in which the size of the gauge 
group responsible for electroweak symmetry breaking is small, because 
the corrections to the electroweak observables become small there. 
We give an estimate for these corrections and find that they are in 
fact phenomenologically acceptable if the size of the gauge group is 
sufficiently small.  Unfortunately, we find that this is the region 
where the theory looses its weakly coupled description, which prevents 
us to make a precise comparison with experiment.  It also implies 
that we have to take into account stringy effects to construct a fully 
well-defined and ultraviolet (UV) completed theory.  However, given 
the presence of an effective field theoretic model and a freedom of 
taking a certain limit, it does not seem so implausible to expect that 
this type of theories does in fact exist.  The experimental signatures 
of such theories are quite distinct.  There is essentially no new state 
appearing below a few TeV ($\approx 2$--$3~{\rm TeV}$) other than the 
standard model gauge bosons and quarks and leptons; in particular, 
there is no Higgs boson.  We then see new states, most of which are 
associated with string states, at the scale of a few TeV.  These states 
arise, in the 4D picture, from the non-trivial dynamics of the new 
strong gauge interaction.  The unitarity of the theory is cured by 
these states and the tail of that physics, which may also be seen 
in scattering experiments at somewhat lower energies than their 
actual masses. 

The organization of the paper is as follows.  In the next section we 
give a general discussion on theories where the electroweak symmetry 
is broken ``dynamically''.  We argue that conventional technicolor-type 
theories and extra dimensional theories with boundary condition 
electroweak symmetry breaking are related in a certain way in the space 
of the gauge coupling, and we elucidate how the electroweak corrections 
in these theories have some similarities.  In section~\ref{sec:model} 
we present the model we study, constructed in the truncated 5D AdS space. 
Electroweak corrections are studied in section~\ref{sec:structure}, 
where we calculate the electroweak oblique parameters and compare 
with experiment.  We discuss two possible scenarios which can be 
phenomenologically viable.  In section~\ref{sec:top} we study the 
top quark sector and its related phenomenology.  Flavor violation 
is also studied there.  Conclusions and discussion are given in 
section~\ref{sec:concl}.

\section{Holography and Electroweak Symmetry Breaking}
\label{sec:holo}

In this paper we mostly study theories formulated in the truncated 
5D AdS space.  This type of theories can provide an understanding 
of a large hierarchy between the Planck and the electroweak scales 
through the AdS warp factor~\cite{Randall:1999ee}.  Before presenting 
an explicit model and going into the detailed calculation, however, 
we here start by some general discussion on theories of ``dynamical'' 
electroweak symmetry breaking.  These include conventional 
technicolor~\cite{Weinberg:gm,Dimopoulos:1979es} and walking 
technicolor~\cite{Holdom:1981rm} theories, as well as theories based 
on extra dimensions such as the ones on flat~\cite{Csaki:2003dt,%
Barbieri:2003pr} or warped~\cite{Csaki:2003zu,Nomura:2003du} geometries. 
We will see that these theories are related in a certain way in the space 
of the gauge coupling and the size of the gauge group.

In order to break the electroweak symmetry dynamically in the IR, we 
need some gauge interaction $G$ that becomes non-perturbative at low 
energies.  We denote the coupling and the size (the number of ``colors'') 
of this gauge interaction as $\tilde{g}$ and $N$, respectively.  In general 
4D theories, the coupling $\tilde{g}$ runs with energy.  Suppose now that 
$G$ is a usual asymptotically-free gauge interaction.  In this case the 
theory is weakly coupled at the UV: the parameter $\L \equiv \tilde{g}^2 
N/16\pi^2$, which is the loop expansion parameter of the gauge theory, 
satisfies $\L \ll 1$. In the IR the parameter $\L$ evolves to larger 
values, and at some scale becomes $\L \simeq 1$, where the theory 
exhibits non-trivial dynamical phenomena such as chiral symmetry 
breaking.  Then, if some fields of the $G$ sector are charged under 
$SU(2)_L \times U(1)_Y$, the electroweak symmetry can be broken at 
this scale. This is the situation in conventional technicolor 
theories~\cite{Weinberg:gm,Dimopoulos:1979es}. Alternatively, $\L$ 
could approach to some constant value close to but somewhat smaller 
than $1$ at the UV, instead of $\L \rightarrow 0$ (or decrease only 
very slowly near $\Lambda$).  Such is the case in walking technicolor 
theories~\cite{Holdom:1981rm}.

Now, let us consider very different possibilities.  At the electroweak 
scale the interaction $G$ induces non-trivial dynamical phenomena. 
Is it then possible for $\L$ to take larger values than $1$ at the UV, 
instead of smaller values?  At first sight, this does not make sense, 
because the loop expansion parameter of the theory is larger than unity 
at the UV --- in fact, the description based on the 4D gauge theory can 
completely break down.  However, in the parameter region $\L \simgt 1$, 
another (sometimes weakly coupled) description of the theory could emerge. 
Suppose we take the limit $N \gg 1$, keeping $\L$ fixed to some value 
larger than unity.  In this case, the loop diagrams are sorted by the 
topology of the graphs and we find that diagrams with different topologies 
correspond to ones having different powers of $N$, allowing us to expand 
the theory in powers of $1/N$~\cite{'tHooft:1973jz}.  Since this expansion 
is reminiscent to the loop expansion by the topology of the world-sheet 
in string theory, the gravitational description of the theory emerges. 
This dual gravitational theory possesses spacetime dimensions larger 
than four, as required by string theory.  The parameter $1/N$ plays a role 
of the (string) coupling constant, while the value of $\L$ turns out to be 
a measure of the importance of string corrections~\cite{Maldacena:1997re}. 
Therefore, for sufficiently large values of $N$ the gravitational 
description is weakly coupled.  For $\L \gg 1$ the string corrections 
are small, corresponding to the region where the curvature scale of the 
gravitational background is much smaller than the string scale, while 
for $\L \simeq 1$ the string effects are important. 

We can now consider the following scenario.  At the UV the theory has 
$\L$ which is close to but somewhat larger than $1$.  The coupling $\L$ 
is almost constant (conformal) over a wide energy interval, but at some 
IR scale this conformality breaks down, triggering non-trivial gauge 
dynamics. In particular, it induces chiral symmetry breaking and 
consequently breaks the electroweak symmetry.  In the dual gravitational 
description, this theory will look like a 5D theory on AdS, as suggested 
by the isomorphism between the 4D conformal group and the isometry of 5D 
AdS space~\cite{Maldacena:1997re}.  The scale of AdS curvature is smaller 
than the string scale if $\L$ is larger than unity.  The non-trivial 
IR dynamics is then represented by the presence of a boundary in the 
spacetime, beyond that point the gravitational description disappears, 
i.e. the gauge theory ``confines''.  We can thus conjecture that the 
theory has a 5D description, compactified on a warped $S^1/Z_2$ orbifold 
with the boundary condition on the IR brane breaking the electroweak 
symmetry.%
\footnote{Strictly speaking, this will be the case only for certain 
special 4D gauge theories.  For instance, weakly-coupled 5D theories 
have a feature that the resonances having spin larger than two are much 
heavier than the others, which is not a property of generic large $N$ 
gauge theories.}
This type of theories have been considered in~\cite{Csaki:2003zu,%
Nomura:2003du} and will be described in the next section.  The 
simple relation between the 4D and 5D theories holds only at energies 
lower than the AdS curvature scale $k$, beyond which the models 
of~\cite{Csaki:2003zu,Nomura:2003du}, for instance, appear intrinsically 
five dimensional.  This scale, however, is much higher than the electroweak 
scale (close to the Planck scale).  An interesting point is that these 
theories allow a large energy interval above the electroweak scale, in 
which the gravitational description does not break down.  This is due to 
the large warp factor of AdS, or the near conformal nature of the theory. 

Models on 5D flat space, such as the ones considered in~\cite{Csaki:2003dt,%
Barbieri:2003pr}, can be obtained from models on AdS by taking the limit 
that the AdS curvature scale is small, $k \rightarrow 0$.  In this case, 
however, the scale where the simple 4D/5D correspondence breaks down 
becomes close to the electroweak scale.  In fact, there is no energy 
interval where the field theoretic correspondence works, and the theory 
appears five dimensional right above the scale of dynamical electroweak 
symmetry breaking.  Some 4D interpretation of the theory, however, may 
be possible for some purposes that do not involve physics much above 
the electroweak scale.

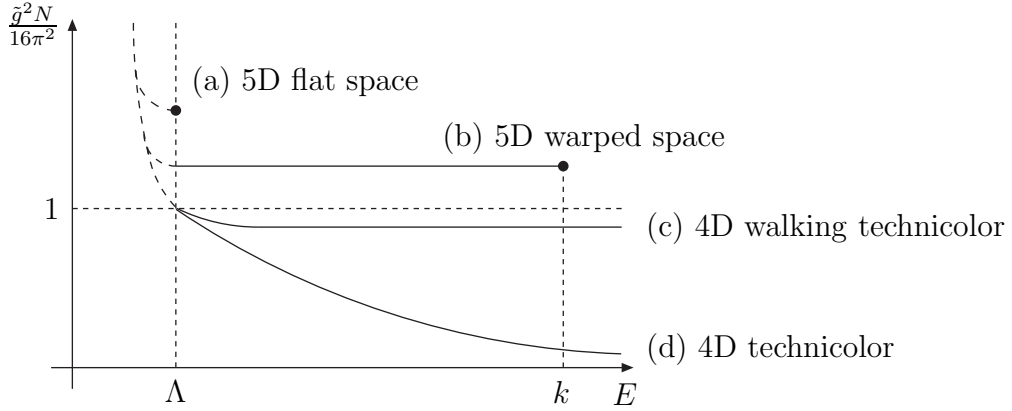
\begin{figure}[t]
\begin{center}
\begin{picture}(300,160)(-15,-20)
  \DashCArc(58,88)(40,187,223){3}  \DashCArc(233,130)(220,180,193){3} 
  \DashCArc(29,88)(12,180,270){3}  \DashCArc(29,112)(15,180,270){3}
  \Text(207,7)[l]{(d) 4D technicolor}
  \CArc(208,325)(320,236,268)
  \Text(207,53)[l]{(c) 4D walking technicolor}
  \CArc(60,123)(70,244,270) \Line(60,53)(197,53)
  \Text(130,84)[bl]{(b) 5D warped space}
  \Line(29,76)(175,76)  \DashLine(175,76)(175,0){2}
  \Vertex(175,76){2}  \Text(175,-5)[t]{$k$}
  \Text(35,105)[bl]{(a) 5D flat space}
  \Vertex(29,97){2}
  \DashLine(-10,60)(197,60){2} \Text(-15,60)[r]{$1$}
  \DashLine(29,0)(29,130){2} \Text(29,-5)[t]{$\Lambda$}
  \LongArrow(-10,-8)(-10,130) \Text(-15,130)[r]{$\frac{\tilde{g}^2 N}{16\pi^2}$}
  \LongArrow(-18,0)(200,0) \Text(199,-6)[t]{$E$}
\end{picture}
\caption{Schematic description for the evolution of the coupling parameter 
 $\L \equiv \tilde{g}^2 N/16\pi^2$ in various theories of dynamical 
 electroweak symmetry breaking.  The behaviors of (a), (b), (c), and (d) 
 represent those of 5D flat space, 5D warped space, walking technicolor 
 and technicolor theories, respectively.} 
\label{fig:holo}
\end{center}
\end{figure}

The schematic behavior of each of these four types of theories is 
depicted in Fig.~\ref{fig:holo} as a function of the energy $E$. 
The theories (a), (b), (c), and (d) correspond, respectively, to 5D flat 
space, 5D warped space, walking technicolor, and technicolor theories. 
The parameter $\Lambda$ represents the scale where $G$ exhibits 
non-trivial IR dynamics, especially chiral symmetry breaking, which 
is roughly the mass of the first resonance state and not much different 
from the electroweak scale.  For $\L \gg 1$, the predictions of the theory 
depend quite little on the value of $\L$, because the theory admits an 
expansion in $1/\L$, with higher order terms corresponding to stringy 
corrections, so that physical quantities are almost determined by the first 
term in the expansion, which is independent of $\L$.  This implies that 
electroweak oblique corrections, whose contributions come from $E \approx 
\Lambda$, can have similar structure in the four types of theories, all of 
which have $\L \simgt 1$ at the scale $\Lambda$.  In particular, the size 
of the corrections essentially depends only on a single parameter $N$. 
However, there is an important difference between the theories of the 
types (a),(b) and (c),(d).  In theories (a) and (b) the electroweak 
oblique parameters are calculable for $N \gg 1$, as the higher order 
terms in the double expansions in $1/N$ and $1/\L$ are both negligible, 
while theories (c) and (d) do not have such calculational powers because 
$\L \simeq 1$ at $E \simeq \Lambda$.  This calculability, however, is lost 
in theories (a) and (b) when we make $N$ or $\L$ smaller.  For the physics 
occurring above $\Lambda$, such as fermion mass generation, the physical 
pictures could be quite different in different types of theories. 

In the following sections we focus on theories in the truncated 5D 
AdS space (the type (b) theories in Fig.~\ref{fig:holo}). In particular, 
we consider electroweak radiative corrections in these theories in 
section~\ref{sec:structure}.  As explained, their structure is expected 
to be similar to the one in technicolor.  Our analysis based on the 
large $N$ expansion will explicitly demonstrate that they are, in fact, 
very similar.%
\footnote{A numerical coincidence between technicolor and 5D flat 
theories found in~\cite{Barbieri:2003pr} also suggests that this 
similarity could be a very precise one.}
While we perform the analysis for the case of warped space theories, 
our qualitative results given in section~\ref{sec:structure} also 
apply to the case of flat space theories by replacing the AdS curvature 
$k$ by the size of the flat extra dimension: $k \rightarrow 1/\pi R$.

\section{The Model}
\label{sec:model}

In this section we review the model constructed in Ref.~\cite{Nomura:2003du}, 
which we will explicitly work on in the rest of the paper.  This model 
reproduces many successful features of the standard model, including fermion 
mass generation and suppression of FCNCs.  The gauge sector of the model is 
quite generic and contains, for example, that of~\cite{Csaki:2003zu} as 
a special point in the parameter space.

The theory is formulated in the 5D warped space with the extra dimension 
compactified on $S^1/Z_2$.  The metric is given by
\begin{equation}
  d s^2 \equiv G_{MN} dx^M dx^N 
    = e^{-2k|y|} \eta_{\mu\nu} dx^\mu dx^\nu + dy^2,
\label{eq:metric}
\end{equation}
where $y$ is the coordinate of the fifth dimension and the physical 
space is taken to be $0 \leq y \leq \pi R$.  We take the AdS curvature 
scale $k$ to be around the 4D Planck scale, and we choose the radius 
$R$ to be $kR \sim 10$.  The scale of the IR brane $k'$ (denoted by 
$T$ in~\cite{Nomura:2003du}) is then given by $k' \equiv k e^{-\pi kR} 
\sim {\rm TeV}$~\cite{Randall:1999ee}.  The fundamental (cutoff) scale 
of the theory is denoted by $M_*$, which is taken to be $M_* \simgt k$, 
and the IR cutoff scale is defined by $M_*' \equiv M_* e^{-\pi kR}$.

The bulk gauge group is $SU(3)_C \times SU(2)_L \times SU(2)_R \times 
U(1)_X$.  It is broken by boundary conditions imposed at the $y=0$ 
brane (the Planck brane):
\begin{equation}
  \partial_y A^{La}_\mu = 0, \quad 
  A^{R1,2}_\mu = 0, \quad
  \partial_y \Bigl( \frac{1}{g_R^2}A^{R3}_\mu 
    + \frac{1}{g_X^2}A^X_\mu \Bigr)=0, \quad
  A^{R3}_\mu - A^X_\mu=0,
\label{eq:bc-0}
\end{equation}
and at the $y=\pi R$ brane (the TeV brane):
\begin{equation}
  \partial_y \Bigl( \frac{1}{g_L^2}A^{La}_\mu 
    + \frac{1}{g_R^2}A^{Ra}_\mu \Bigr) = 0, \quad 
  A^{La}_\mu - A^{Ra}_\mu = 0, \quad 
  \partial_y A^X_\mu = 0,
\label{eq:bc-pi}
\end{equation}
with the $A_5$'s obeying Dirichlet (Neumann) boundary conditions 
if the corresponding $A_\mu$'s obey Neumann (Dirichlet) boundary 
conditions~\cite{Csaki:2003zu,Csaki:2003dt} (these boundary conditions 
are slightly modified when brane-localized gauge kinetic terms are 
introduced at $y=0$ and $y=\pi R$).  At the Planck brane, the bulk 
gauge group is broken to the standard-model gauge group $SU(3)_C \times 
SU(2)_L \times U(1)_Y$, where $U(1)_Y$ is a linear combination of 
$U(1)_X$ and the $T_3$ direction of $SU(2)_R$, while at the TeV brane, 
$SU(2)_L \times SU(2)_R$ is broken to the $SU(2)$ diagonal subgroup. 
Combining the breaking at the both branes, the unbroken gauge group 
at low energies becomes $SU(3)_C \times U(1)_{\rm EM}$, where 
$U(1)_{\rm EM}$ refers to electromagnetism.  In particular, the 
electroweak symmetry is broken by the boundary conditions at 
the TeV brane.

We can view the above boundary conditions as the limiting case of 
the following brane Higgs breaking.  We introduce a scalar field 
$\Sigma({\bf 1}, {\bf 1}, {\bf 2}, 1/2)$ on the Planck brane and 
$H({\bf 1}, {\bf 2}, {\bf 2}^*, 0)$ on the TeV brane, where the 
numbers in the parentheses represent gauge quantum numbers under 
$SU(3)_C \times SU(2)_L \times SU(2)_R \times U(1)_X$. Now, suppose 
that these fields have VEVs:
\begin{equation}
  \langle \Sigma \rangle = \pmatrix{0 \cr v_\Sigma},
\qquad
  \langle H \rangle = \pmatrix{v_H & 0 \cr 0 & v_H}.
\label{eq:b-higgs}
\end{equation}
We then find that for $v_\Sigma, v_H \rightarrow \infty$, the 
phenomenology of this Higgs-breaking theory becomes identical to 
that of the boundary-condition breaking theory~\cite{Nomura:2001mf}. 
In particular, the physical Higgs bosons arising from $\Sigma$ and $H$ 
decouple for large $v_\Sigma$ and $v_H$, so that there is no scalar 
particle remaining in the spectrum in this limit. 

The kinetic terms for the gauge fields are given by
\begin{eqnarray}
  && {\cal S} = \int\!d^4x \int\!dy \Biggl[ \sqrt{-G}
    \Biggl\{ -\frac{1}{4g_L^2} g^{MP}g^{NQ} \sum_{a=1}^{3} 
      F^{La}_{MN} F^{La}_{PQ} 
  -\frac{1}{4g_R^2} g^{MP}g^{NQ} \sum_{a=1}^{3} 
      F^{Ra}_{MN} F^{Ra}_{PQ} 
\nonumber\\
  && -\frac{1}{4g_X^2} g^{MP}g^{NQ} F^{X}_{MN} F^{X}_{PQ} \Biggr\}
  + \delta(y) \Biggl\{ 
  -\frac{1}{4\tilde{g}_L^2} 
    \sum_{a=1}^{3} F^{La}_{\mu\nu} F^{La}_{\mu\nu}
  -\frac{1}{16\tilde{g}_Y^2} 
    \bigl( F^{R3}_{\mu\nu} + F^X_{\mu\nu} \bigr) 
    \bigl( F^{R3}_{\mu\nu} + F^X_{\mu\nu} \bigr) \Biggr\} \Biggr],
\label{eq:gauge-kinetic}
\end{eqnarray}
where $F^{La}_{MN}$, $F^{Ra}_{MN}$ and $F^{X}_{MN}$ are the 
field-strength tensors for $SU(2)_L$, $SU(2)_R$ and $U(1)_X$, 
and $g_L$, $g_R$ and $g_X$ are the 5D gauge couplings having 
mass dimensions $-1/2$; $a$ is the indices for the adjoint 
representation of $SU(2)$.  Here, we have included Planck-brane 
localized gauge kinetic terms, which are radiatively generated and 
generically have coefficients of order $(b/8\pi^2)\ln(k/k') \sim 1$. 
TeV-brane localized gauge kinetic terms are considered in the next 
section.  We have omitted the gauge kinetic terms for $SU(3)_C$ in the 
above expression, since they are irrelevant for our discussion below.

The quarks and leptons are introduced in the bulk with the 
representations: 
\begin{eqnarray}
  && q({\bf 3}, {\bf 2}, {\bf 1}, 1/6), \qquad
  \bar{u} = \psi_{\bar{u}}({\bf 3}^*, {\bf 1}, 
      {\bf 2}, -1/6)|_{T^R_3 = -1/2}, \qquad
  \bar{d} = \psi_{\bar{d}}({\bf 3}^*, {\bf 1}, 
      {\bf 2}, -1/6)|_{T^R_3 = 1/2}, 
\nonumber\\
  && l({\bf 1}, {\bf 2}, {\bf 1}, -1/2), \qquad
  \bar{e} = \psi_{\bar{e}}({\bf 1}, {\bf 1}, 
      {\bf 2}, 1/2)|_{T^R_3 = 1/2}, \qquad
  [\bar{n} = \psi_{\bar{n}}({\bf 1}, {\bf 1}, 
      {\bf 2}, 1/2)|_{T^R_3 = -1/2}],
\label{eq:fermion-rep}
\end{eqnarray}
where $q,\bar{u},\bar{d},l,\bar{e}$ and $\bar{n}$ are Dirac fermions 
and the numbers in the parentheses represent gauge quantum numbers 
under $SU(3)_C \times SU(2)_L \times SU(2)_R \times U(1)_X$; $T^R_3 
= \pm 1/2$ represents the $T_3 = \pm 1/2$ component of the $SU(2)_R$ 
doublet.  With the extra dimension compactified on $S^1/Z_2$, we can 
arrange the boundary conditions such that only the left-handed components 
of $q,\psi_{\bar{u}},\psi_{\bar{d}},l$ and $\psi_{\bar{e}}$ possess 
zero modes (also $\psi_{\bar{n}}$ if we introduce them to induce small 
neutrino masses through the see-saw mechanism; see~\cite{Nomura:2003du,%
Goldberger:2002pc}). Moreover, introducing the Planck-brane localized 
left-handed fermions $\psi'_{\bar{u}}({\bf 3}, {\bf 1}, {\bf 1}, -1/3)$, 
$\psi'_{\bar{d}}({\bf 3}, {\bf 1}, {\bf 1}, 2/3)$ and 
$\psi'_{\bar{e}}({\bf 1}, {\bf 1}, {\bf 1}, 0)$, and the couplings 
$\delta(y) [\psi_{\bar{u}} \psi'_{\bar{u}} \Sigma + \psi_{\bar{d}} 
\psi'_{\bar{d}} \Sigma^\dagger + \psi_{\bar{e}} \psi'_{\bar{e}} 
\Sigma^\dagger]$, we can make the unwanted zero modes from the 
$T^R_3 = 1/2$ component of $\psi_{\bar{u}}$ and the $T^R_3 = -1/2$ 
components of $\psi_{\bar{d}}$ and $\psi_{\bar{e}}$ heavy to get 
masses of order $k$.%
\footnote{This also makes the theory anomaly free together with the 
introduction of appropriate Chern-Simons terms~\cite{Arkani-Hamed:2001is}.}
The low-energy matter content is then precisely that of the 
standard-model quarks and leptons, which arise as the zero modes 
of $q,\bar{u},\bar{d},l$ and $\bar{e}$.  The quark and lepton 
masses arise from the couplings introduced on the TeV brane
\begin{equation}
  {\cal S} = \int\!d^4x \int\!dy \: \delta(y-\pi R) \sqrt{-g_{\rm ind}} 
    \Biggl[ y_u q \psi_{\bar{u}} H + y_d q \psi_{\bar{d}} H 
    + y_e l \psi_{\bar{e}} H + {\rm h.c.} \Biggr],
\label{eq:yukawa}
\end{equation}
where we have suppressed the generation index.  As the up-type quark, 
down-type quark, and charged lepton masses arise from three independent 
couplings, $y_u$, $y_d$ and $y_e$, there are no unwanted relations 
among them coming from $SU(2)_R$. 

The wavefunction profiles for the zero modes of the quark and 
lepton fields are controlled by the 5D bulk mass parameters for 
these fields, which we parameterize as ${\cal L}_{\rm 5D} \supset 
-c k \bar{\Psi} \Psi$ where $\Psi$ represents generic 5D (Dirac) 
fermions.  For $c > 1/2$ ($c < 1/2$) the wavefunction for the 
left-handed zero-mode fermion is localized to the Planck (TeV) brane. 
We take parameters $c$ to be larger than $1/2$, at least for the 
first-two generation fermions.  This makes the non-universality 
of the $W$- and $Z$-boson couplings to these fields very small 
so that the theory is phenomenologically viable~\cite{Nomura:2003du}, 
and could also provide a partial understanding of the flavor structure 
of the fermion mass matrices and suppression of the flavor violation 
arising from the TeV-brane operators~\cite{Gherghetta:2000qt}. 
The $c$ parameters for the third generation fermions are the theme 
of section~\ref{sec:top}, where the non-universality of the fermion 
gauge couplings is also discussed further.

We can now calculate the masses and couplings of the electroweak gauge 
bosons, $W$, $Z$ and $\gamma$.  Assuming $1/g_L^2 \sim 1/g_R^2 \sim 
1/g_X^2 \sim 1/\pi R$ and $\tilde{g}_L \sim \tilde{g}_Y \sim 1$, we 
find that these masses and couplings take exactly the same form as that 
of the standard model, at the leading order in $1/\pi kR$ and in $k'/k$. 
Denoting the standard-model $SU(2)_L$ and $U(1)_Y$ couplings as $g$ 
and $g'$ and the Higgs-field VEV as $v \simeq 175~{\rm GeV}$, the 
correspondence between the two theories are given by~\cite{Nomura:2003du}
\begin{equation}
  \frac{1}{g^2} = \frac{\pi R}{g_L^2} + \frac{1}{\tilde{g}_L^2}, \qquad
  \frac{1}{g'^2} = \frac{\pi R}{g_R^2} + \frac{\pi R}{g_X^2} 
    + \frac{1}{\tilde{g}_Y^2}, \qquad
  v^2 = \frac{4\, k'^2}{(g_L^2 + g_R^2)k}.
\label{eq:corresp}
\end{equation}
Thus, for given values of the brane couplings, $\tilde{g}_L$ and 
$\tilde{g}_Y$, we have two relations on the three bulk gauge couplings, 
$g_L$, $g_R$ and $g_X$, so that we can calculate various quantities 
in terms of a single free parameter, which we take to be $g_R^2/g_L^2$ 
(an introduction of TeV-brane gauge kinetic terms will give a few extra 
parameters).   For instance, the value of $k'$ is determined by the last 
equation of Eqs.~(\ref{eq:corresp}), which in turn gives the masses of 
the Kaluza-Klein (KK) gauge bosons
\begin{equation}
  m_n \simeq \frac{\pi}{2} \Bigl( n + \frac{1}{2} \Bigr) k',
\label{eq:KK-gauge}
\end{equation}
where $n=1,2,3,4\cdots$ for the $W$ and $Z$ towers and $n=1,3,5,\cdots$ 
for the $\gamma$ and gluon towers~\cite{Nomura:2003du}. 

What are values of the Planck-brane gauge couplings?  
In Eqs.~(\ref{eq:corresp}) we have to use values of $\tilde{g}_L$ 
and $\tilde{g}_Y$ appropriately normalized at the scale around TeV. 
Since the running of the Planck-brane gauge couplings is determined 
by the zero modes (elementary fields in the dual picture) of the 
theory~\cite{Goldberger:2002cz}, we can write $1/\tilde{g}_L^2$ 
and $1/\tilde{g}_Y^2$ as
\begin{equation}
  \frac{1}{\tilde{g}_L^2} = \frac{1}{\tilde{g}_{L,0}^2}
    + \frac{b_L}{8\pi^2}\ln\Bigl(\frac{k}{k'}\Bigr),
\qquad
  \frac{1}{\tilde{g}_Y^2} = \frac{1}{\tilde{g}_{Y,0}^2}
    + \frac{b_Y}{8\pi^2}\ln\Bigl(\frac{k}{k'}\Bigr),
\label{eq:brane-kin-TeV}
\end{equation}
where $(b_L,b_Y) \simeq (-10/3,20/3)$ in the present theory. 
The couplings $\tilde{g}_{L,0}$ and $\tilde{g}_{Y,0}$ represent the 
running couplings of the elementary $SU(2)_L$ and $U(1)_Y$ gauge bosons 
at the scale $k$.  In general, these couplings are free parameters of 
the theory and cannot be calculated in the effective theory.  

One natural possibility is to assume that the elementary sector 
of the theory is strongly coupled at the scale $k$, in which case the 
bare parameters, $\tilde{g}_{L,0}$ and $\tilde{g}_{Y,0}$, are estimated 
to be $1/\tilde{g}_{L,0}^2 \sim 1/\tilde{g}_{Y,0}^2 \sim 1/16\pi^2$ 
through naive dimensional analysis (NDA)~\cite{Manohar:1983md}.  This is 
the case considered in~\cite{Nomura:2003du} (and in~\cite{Csaki:2003zu}), 
leading to the situation that the free parameters in the gauge sector of 
the theory are effectively only $g_R^2/g_L^2$ and $M_*/k$ (and TeV-brane 
localized kinetic terms).  Note, however, that, contrary to the flat 
space case, in warped space theories the strong-coupling requirement 
for the Planck-brane operators is independent from the strong-coupling 
requirement for the bulk and TeV-brane operators.  For instance, it is 
completely natural to assume that all the bulk and TeV-brane operators 
scale according to NDA while couplings at the Planck brane are weak. 
(In the dual 4D picture, this is equivalent to requiring that the $G$ 
sector does not contain any small or large dimensionless parameter other 
than the size of the gauge group, while the elementary sector is weakly 
coupled.)  In this case the observed 4D gauge couplings are almost entirely 
given by the Planck-brane couplings, $1/g^2 \simeq 1/\tilde{g}_L^2$ 
and $1/g'^2 \simeq 1/\tilde{g}_Y^2$, and the bulk gauge couplings 
take the values determined by NDA, $1/g_L^2 \sim 1/g_R^2 \sim 1/g_X^2 
\sim M_*/16\pi^3$ (a similar scenario has been considered in flat 
space in~\cite{Barbieri:2003pr}).%
\footnote{Here we do not bother the difference between the 4D loop 
factor $16\pi^2$ and the 5D loop factor $24\pi^3$ too much, and adopt 
a somewhat ``conservative'' estimate using $16\pi^3$.  This gives a 
strong coupling value for the 4D gauge coupling, $g_{\rm 4D} \simeq 4\pi$, 
when the IR cutoff $M'_*$ is lowered to the mass of the first KK resonance, 
$M'_* \simeq \pi k'$ (see the last paper in~\cite{Manohar:1983md}). 
We also do not include group theoretical factors for NDA because the 
bulk gauge groups (i.e. $SU(2)_L, SU(2)_R$ and $U(1)_X$, {\it not} $G$) 
are small.}
In either case the largest value of $k'$ is determined by the 
parameter $M_*/(\pi k)$ as $k'|_{\rm max} \approx (1 \sim 
1.5~{\rm TeV})(\pi k/M_*)^{1/2}$, which is translated into the 
maximum value for the lowest KK gauge boson mass $m_1|_{\rm max} 
\approx (2.5 \sim 3.5~{\rm TeV})(\pi k/M_*)^{1/2}$.  If we require 
that the theory admits a weakly coupled description, e.g. $M_*/(\pi k) 
\simgt 3$, this gives $k'|_{\rm max} \approx (600 \sim 900)~~{\rm GeV}$ 
and $m_1|_{\rm max} \approx (1.5 \sim 2)~{\rm TeV}$.

In the next section we consider corrections to the electroweak 
observables in the present theory.  Our analysis applies regardless 
of the values of $\tilde{g}_{L,0}$ and $\tilde{g}_{Y,0}$, and thus 
to either of the two cases described above.

\section{Electroweak Corrections from the Gauge Sector}
\label{sec:structure}

In this section we discuss the structure of the corrections to  
electroweak observables in the theory presented in the previous section.  
We concentrate here on the corrections from the pure gauge sector and 
leave those from the matter sector to the next section.  We consider the 
electroweak oblique parameters $S$, $T$ and $U$~\cite{Peskin:1990zt} and 
give estimates for them.  We elucidate how the 4D dual picture provides 
a qualitative understanding of the structure of these corrections. 
We then calculate the leading corrections in the 5D picture and compare 
them with the results of 4D considerations, which quantitatively 
demonstrates the duality between the two theories.  We find that, if we 
stick to the presence of a weakly coupled gravitational description of the 
theory, the model gives somewhat larger (positive) values of $S$ than those 
allowed by precision electroweak measurements.  There are essentially two 
ways out of this unpleasant situation.  One is to extend the model such 
that it has a sector giving a negative value of $S$ canceling the positive 
contribution.  The other is to give up the weakly coupled description 
of the theory.  In particular, we argue that once we depart from the 
weakly coupled description, the theory could avoid constraints from 
the precision measurements. 

Some of the discussions in this section overlap with those 
in~\cite{Barbieri:2003pr}, which explicitly considers these issues 
in a flat space model with some discussions on general gravitational 
backgrounds.  Our explicit result for the warped space model agrees 
with the expectation given in~\cite{Barbieri:2003pr}.

\subsection{Structure of electroweak corrections}
\label{subsec:structure-1}

We start with the 4D dual picture of the theory.  As discussed in 
section~\ref{sec:holo}, we can relate the theory described in 
the previous section to a purely 4D theory through the AdS/CFT 
correspondence~\cite{Maldacena:1997re,Arkani-Hamed:2000ds}. 
In this 4D dual picture, the theory below $k \sim M_{\rm pl}$ 
contains a gauge interaction with the group $G$, whose coupling evolves 
very slowly over a wide energy interval below $k$.  This $G$ gauge 
sector possesses a global $SU(3)_C \times SU(2)_L \times SU(2)_R \times 
U(1)_X$ symmetry whose $SU(3)_C \times SU(2)_L \times U(1)_Y$ subgroup 
is gauged, where $U(1)_Y$ is a linear combination of $U(1)_X$ and the 
$T_3$ direction of $SU(2)_R$.  Therefore, the theory in this energy 
interval appears as $SU(3)_C \times SU(2)_L \times U(1)_Y \times G$ 
gauge theory with the quarks and leptons transforming under $SU(3)_C 
\times SU(2)_L \times U(1)_Y$.  At the TeV scale the gauge interaction 
of $G$ exhibits non-trivial IR phenomena, producing resonances of masses 
of order TeV.  These resonances have a tower structure.  In particular, 
there are towers of spin-$1$ fields which have the quantum numbers 
of $W$, $Z$ and $\gamma$.  These towers then mix with the elementary 
gauge bosons of the weakly gauged $SU(2)_L$ and $U(1)_Y$ groups. 
The resulting spectrum consists of towers of gauge bosons with the 
quantum numbers of $W$ and $Z$, whose lowest states are massive and 
identified as the standard-model $W$ and $Z$ bosons, and a tower of 
$U(1)$ gauge bosons, whose lowest mode is massless and identified 
with the photon.  These towers of mass eigenstates are dual to the 
$W$, $Z$ and $\gamma$ KK towers in the 5D picture.  The electroweak 
gauge group $SU(2)_L \times U(1)_Y$ is dynamically broken and the 
masses of the $W$ and $Z$ bosons and the quarks and leptons are 
generated.

\begin{figure}[t]
\begin{center}
\begin{picture}(100,55)(150,150)
  \GOval(200,180)(25,30)(0){0.8}
  \Photon(120,180)(170,180){3}{5}
  \Photon(280,180)(230,180){3}{5}
  \Text(130,193)[b]{$W_\mu$} \Text(270,193)[b]{$W_\mu$}
\end{picture}
\caption{The diagram contributing to the $W$ boson propagators. 
 Similar diagrams exist with one or two external $W_\mu$'s replaced 
 by $B_\mu$.} 
\label{fig:diag-1}
\end{center}
\end{figure}
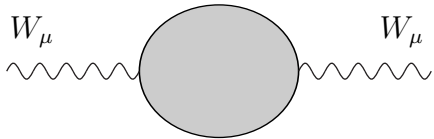

How do the corrections to electroweak gauge boson propagators arise 
in the 4D picture?  We concentrate here on the effect from spin-$1$ 
resonances and leave the consideration of the other effects to the 
next section.  At leading order, the corrections arise from the 
diagrams such as the one given in Fig.~\ref{fig:diag-1}, where the 
gray disk at the center of the diagram represents contributions from 
the strongly interacting $G$ sector. 
We have drawn only the diagram giving corrections to the $SU(2)_L$ 
gauge boson, $W^a_\mu$, but similar diagrams also exist with one or 
two external $W_\mu$'s replaced by the $U(1)_Y$ gauge boson, $B_\mu$. 
These diagrams give a contribution to the $S$ parameter (specifically, 
it arises from the diagram having $W_\mu$ for one external line and 
$B_\mu$ for the other).  To evaluate the contribution, we must know 
what this gray disk actually means.  For a sufficiently large value 
of $N$, this leading order contribution comes from the sum of 
a series of diagrams given in Fig.~\ref{fig:diag-2} (the planar 
diagrams~\cite{'tHooft:1973jz}). 
\begin{figure}[t]
\begin{center} 
\begin{picture}(200,130)(155,85)
  \GOval(100,190)(20,24)(0){0.8} \Text(150,190)[]{$=$}
  \Oval(200,190)(20,24)(0) \Text(220,208)[lb]{$\psi_G$} 
  \Text(240,190)[]{$+$}
  \Oval(280,190)(20,24)(0) \Gluon(280,210)(280,170){2.6}{5}
  \Text(300,208)[lb]{$\psi_G$} \Text(285,190)[l]{$A^G_\mu$} 
  \Text(320,190)[]{$+$}
  \Oval(360,190)(20,24)(0) \Gluon(360,210)(360,190){2.6}{2.5}
  \Gluon(360,190)(345,175){2.6}{2.5} \Gluon(360,190)(375,175){2.6}{2.5}
  \Text(380,208)[lb]{$\psi_G$} \Text(365,194)[l]{$A^G_\mu$} 
  \Text(400,190)[]{$+$} \Text(415,190)[l]{$\cdots$}
  \Text(150,140)[]{$\simeq$} 
  \Text(200,140)[]{\Large $\frac{N}{16\pi^2}$}
  \Text(240,140)[]{$+$} 
  \Text(280,140)[]{\Large $\frac{\tilde{g}^2 N^2}{(16\pi^2)^2}$}
  \Text(320,140)[]{$+$} 
  \Text(360,140)[]{\Large $\frac{\tilde{g}^4 N^3}{(16\pi^2)^3}$}
  \Text(400,140)[]{$+$} \Text(415,140)[l]{$\cdots$}
  \Text(150,100)[]{$=$}
  \Text(175,100)[l]{\Large $\frac{N}{16\pi^2}f(\frac{\tilde{g}^2 N}{16\pi^2})$}
  \Text(280,100)[]{$\sim$} \Text(305,100)[l]{\Large $\frac{N}{16\pi^2}$}
\end{picture}
\caption{The diagrams represented by the gray disk.  Here, $\psi_G$ and 
 $A^G_\mu$ represent matter and gauge fields of the strongly-coupled $G$ 
 sector.  The size of each diagram is also shown.  The contribution from 
 this set of diagrams will be of order $N/16\pi^2$.} 
\label{fig:diag-2}
\end{center}
\end{figure}
In the figure we have given the size of contributions from each 
diagram.  Writing the contribution from all the diagrams in the form 
$(N/16\pi^2) f(\tilde{g}^2 N/16\pi^2)$ where $f(x)$ is some function, we 
expect that the gray disk gives a contribution of order $N/16\pi^2$ in 
Fig.~\ref{fig:diag-1} and thus changes the coefficients of the gauge kinetic 
terms from $1/g^2$ to $1/g^2 + c N/16\pi^2$, where $g$ represents generic 
electroweak gauge couplings and $c$ is a numerical factor of $O(1)$.%
\footnote{For $\L \equiv \tilde{g}^2 N/16\pi^2 \ll 1$, $f(\L)$ has 
an expansion $f(\L) = \sum_0^\infty c_n \L^n$, where $c_n = O(1)$. 
This is the domain where the perturbative 4D gauge theory description 
is appropriate.  For $\L \gg 1$, $f(\L)$ again has an expansion but of 
the form $f(\L) = \sum_0^\infty c'_n \L^{-n}$, where $c'_n = O(1)$. This 
is the region where the theory is well described by a semi-classical 
gravitational theory, with higher order terms in the expansion 
corresponding to corrections from string theory.  In the region 
$\L \simeq 1$, neither description is good and in the absence of 
the explicit string realization of the theory, we can only say that 
$f(\L) = O(1)$. In particular, for $\L \simeq N \simeq 1$, the theory 
does not admit any weakly coupled description and can only be described 
by strongly coupled ($g_s \simeq 1$) string theory.}
This gives a contribution to the $S$ parameter
\begin{equation}
  S \simeq \frac{N}{\pi},
\label{eq:S-estimate}
\end{equation}
since $S$ is given by $S \equiv 16\pi (\Pi'_{33} - \Pi'_{3Q})$, 
where $\Pi'_{XY}$ are the corrections to the gauge kinetic terms defined 
by ${\cal L} = -(1/4)(1/g^2 - \Pi'_{XY}) F^X_{\mu\nu}F^Y_{\mu\nu}$ 
(see Appendix~A for details).%
\footnote{Here we have taken matter in the $G$ sector, $\psi_G$, to be 
in the fundamental representation with the $O(1)$ number of ``flavors''.  
This may not be the case in our actual theory because the near conformality 
of the theory would require a large matter sector.  However, for simplicity 
we will keep presenting our analysis for this simple matter sector, as it 
will give the correct relations between the physical quantities.  We will 
come back to this point at the end of this subsection.}

To derive the value of $N$ in the present theory, we consider the 
$G$ sector as an analogue of the QCD.  The standard analysis in large 
$N$ QCD~\cite{Witten:1979kh} gives the relation between the pion decay 
constant, $f_\pi$, and the mass of the lowest spin-$1$ resonance ($\rho$ 
meson), $m_{\rho}$, as
\begin{equation}
  f_\pi \simeq \frac{\sqrt{N}}{4\pi} m_{\rho}.
\label{eq:fpi-mrho}
\end{equation}
In the present context, the pion decay constant, $f_\pi$, in the $G$ 
sector corresponds to the electroweak scale $v \simeq 175~{\rm GeV}$ 
and the mass of the $\rho$ meson, $m_{\rho}$, to the mass of the 
lowest KK excited gauge boson $m_1 \simeq (3\pi/4)k'$.  Then, rewriting 
$f_\pi$ and $m_\rho$ in Eq.~(\ref{eq:fpi-mrho}) as $v$ and $m_1$ and 
using Eq.~(\ref{eq:corresp}), we obtain the number of ``colors'', 
$N$, for $G$:%
\footnote{Equivalently, the value of $N$ can be determined by the 
following argument.  The diagram of Fig.~\ref{fig:diag-1} gives 
the squared masses for $W$ and $Z$ of order $(N/16\pi^2)m_1^2$, 
in the normalization where the gauge couplings appear in front of 
the gauge kinetic terms.  Since these masses are $v^2$, we obtain 
$N \simeq 16\pi^2 v^2/m_1^2$.}
\begin{equation}
  N \simeq 16\pi^2 \frac{v^2}{m_1^2} 
    \simeq \frac{16\pi^2}{(g_L^2+g_R^2)k}.
\label{eq:def-N}
\end{equation}
Combined with Eq.~(\ref{eq:S-estimate}), this equation tells us 
that the correction to the $S$ parameter becomes smaller if we make 
the first KK states heaver, {\it i.e.} $m_1/v$ larger, which can 
be attained by making either $g_L^2 k$ or $g_R^2 k$ larger. 
As was found in~\cite{Nomura:2003du}, this can be done by making 
$g_R^2/g_L^2$ larger even in the case of $1/\tilde{g}_{L,0}^2 \simeq 
1/\tilde{g}_{Y,0}^2 \simeq 1/16\pi^2$.  An alternative possibility 
is to take all the bulk gauge couplings large, $1/g_L^2 \simeq 
1/g_R^2 \simeq 1/g_X^2 \simeq M_*/16\pi^3$, and to assume that the 4D 
gauge couplings almost entirely come from the Planck-brane couplings.

Here we comment on the range of $N$ we are imagining.  Because 
the 4D gauge couplings $g_{\rm 4D}$ receive contributions from 
the bulk gauge couplings, $g_{\rm 5D}$, we naturally expect that 
$1/g_{\rm 4D}^2 \simgt \pi R/g_{\rm 5D}^2$ (see Eq.~(\ref{eq:corresp}), 
for example).  This relation can be written as $1/g_{\rm 4D}^2 
\simgt \pi k R/g_{\rm 5D}^2 k \simeq (N/16\pi^2) \ln(k/k')$, using 
Eq.~(\ref{eq:def-N}).  Since the observed 4D gauge couplings are 
of order $1$, this gives constraints on $N$: $N \simlt 16\pi^2/\ln(k/k')$.
This relation is understood in the 4D picture as the condition that 
the asymptotically non-free running of the 4D gauge couplings caused 
by the $G$ sector, $(N/16\pi^2) \ln(k/k')$, must not make the low-energy 
values of the gauge couplings, $g_{\rm 4D}$, too small.  In any case, 
with $\ln(k/k') \simeq 30$, we obtain $N \simlt 5$ so that we are 
not considering very large values of $N$ in the present context.

At leading order in $1/N$, represented by the diagram in 
Fig.~\ref{fig:diag-1}, the $T$ and $U$ parameters are not generated. 
This is because the $G$ sector respects the global custodial $SU(2)$ 
symmetry so that just inserting the $G$ dynamics, i.e. the gray disk, 
does not give $T$ or $U$ parameters~\cite{Agashe:2003zs}.  Therefore, 
at this order, the electroweak oblique parameters receive contributions
\begin{equation}
  S = c_S \frac{16\pi}{(g_L^2+g_R^2)k},
\label{eq:contr-s}
\end{equation}
\begin{equation}
  T = U = 0,
\end{equation}
where $c_S$ is a coefficient of order unity. 

\begin{figure}[t]
\begin{center}
\begin{picture}(140,55)(130,160)
  \GOval(200,180)(5,50)(0){0.8}
  \Photon(110,180)(150,180){3}{5}
  \Photon(290,180)(250,180){3}{5}
  \Text(130,193)[b]{$W_\mu$} \Text(270,193)[b]{$W_\mu$}
  \DashLine(200,200)(200,160){5} \Text(200,207)[b]{\large $A$}
\end{picture}
\caption{The diagram representing the mixing between the elementary 
 $W$ boson and the composite $W$ states arising from the dynamics of $G$.
 Cutting the diagrams at $A$ gives the states which have the same quantum 
 numbers and spin as $W$.  Similar diagrams also exist for $W_\mu$ 
 replaced by the gauge boson of $U(1)_Y$, $B_\mu$.}
\label{fig:diag-3}
\end{center}
\end{figure}
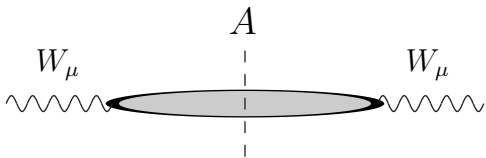

What does this leading order contribution correspond to in the 5D 
picture?  To see this, it is instructive to write Fig.~\ref{fig:diag-1} 
in a slightly deformed way as in Fig.~\ref{fig:diag-3}.
This diagram can be understood as the one in which the $W$ boson is 
transformed to some states made up of the constituents of the $G$ sector, 
and then goes back to $W$.  Making a cut at the center of the figure 
(the dashed line denoted as $A$ in the figure) we find that these states 
are the bound states of $G$ and have the same quantum numbers and spin 
as $W$.  It is then clear that this represents mixings between the 
elementary $W$ boson and the excited $W$ bosons, the spin-$1$ bound 
states of $G$.  Since these excited states correspond to KK states in 
the 5D picture, we learn that the contribution at leading order 
in $1/N$ corresponds to the tree-level contribution in the 5D picture. 
In fact, by solving the masses and wavefunctions for the electroweak 
gauge bosons at tree level in 5D, we find that the contribution to $S$ 
takes the form of Eq.~(\ref{eq:contr-s}).  In the absence of TeV-brane 
kinetic terms, the coefficient $c_S$ is given by $c_S = 3/4$ (more 
detailed discussions including the TeV-brane operators are given in 
the next subsection and in Appendix~A).  An interesting point here is 
that we can calculate the coefficient in Eq.~(\ref{eq:contr-s}), i.e. 
sum up the planar diagrams, and it gives the dominant contribution for 
sufficiently large $N$, i.e. for sufficiently small 5D gauge couplings 
$(g_L^2 + g_R^2)k \ll 16\pi^2$.

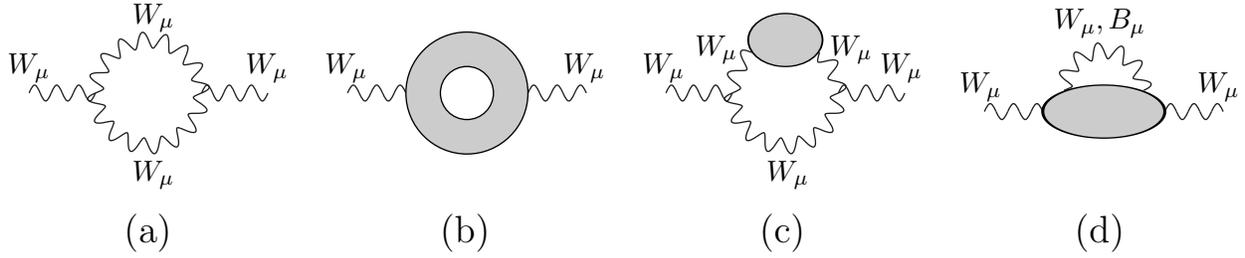
\begin{figure}[t]
\begin{center} 
\begin{picture}(500,100)(5,125)
  \Text(70,145)[t]{\large (a)}
  \Photon(25,190)(50,190){3}{3.13} \Photon(90,190)(115,190){3}{3.13}
  \PhotonArc(70,190)(20,0,180){3}{7.8} \PhotonArc(70,190)(20,180,360){3}{7.8}
  \Text(72,217)[b]{\small $W_\mu$} \Text(72,165)[t]{\small $W_\mu$} 
  \Text(25,198)[b]{\small $W_\mu$} \Text(115,198)[b]{\small $W_\mu$} 
  \Text(190,145)[t]{\large (b)}
  \Photon(145,190)(167,190){3}{2.75} \Photon(213,190)(235,190){3}{2.75}
  \GCirc(190,190){23}{0.8} \GCirc(190,190){10}{1} 
  \Text(145,198)[b]{\small $W_\mu$} \Text(235,198)[b]{\small $W_\mu$} 
  \Text(310,145)[t]{\large (c)}
  \Photon(265,190)(290,190){3}{3.13} \Photon(330,190)(355,190){3}{3.13}
  \PhotonArc(310,190)(20,0,50){3}{2.18} \Text(328,205)[bl]{\small $W_\mu$}
  \PhotonArc(310,190)(20,130,180){3}{2.18} \Text(293,205)[br]{\small $W_\mu$}
  \Text(265,198)[b]{\small $W_\mu$} \Text(355,198)[b]{\small $W_\mu$} 
  \PhotonArc(310,190)(20,180,360){3}{7.8} \Text(312,165)[t]{\small $W_\mu$} 
  \GOval(310,210)(10,14)(0){0.8} 
  \Text(430,145)[t]{\large (d)}
  \Photon(385,183)(407,183){3}{2.73} \Photon(453,183)(475,183){3}{2.73}
  \PhotonArc(430,191)(15,0,180){3}{5.9} \GOval(430,183)(10,23)(0){0.8}
  \Text(430,215)[b]{\small $W_\mu, B_\mu$}
  \Text(385,191)[b]{\small $W_\mu$} \Text(475,191)[b]{\small $W_\mu$} 
\end{picture}
\caption{The diagrams with an elementary loop~(a), with the $G$ effect 
 at the next-to-leading order in $1/N$~(b), and with an additional 
 loop of elementary fields on top of the leading $G$ effect~(c,d).  
 For (a) and (c), similar diagrams using gauge 4-point vertices exist. 
 For (b) and (d), there are also similar diagrams with $B_\mu$ on some 
 of external lines.}
\label{fig:diag-4}
\end{center}
\end{figure}

Before comparing with experiment, we discuss what happens at 
the next order.  The next order contributions come at one loop in 
5D, which corresponds to four different types of diagrams as shown in 
Fig.~\ref{fig:diag-4} (and diagrams with more insertions of the gray 
disk).  The first one (Fig.~\ref{fig:diag-4}a) is the loop of the 
elementary gauge bosons, the second one (Fig.~\ref{fig:diag-4}b) 
represents the diagram at the next-to-leading order in $1/N$, and 
the third and fourth ones (Fig.~\ref{fig:diag-4}c,d) represent 
the ones with an additional loop of elementary fields to that of 
Fig.~\ref{fig:diag-1}. 
The first diagram does not pick up effects of electroweak symmetry 
breaking so that it does not give contributions to $S$, $T$ or $U$.
The third diagram (Fig.~\ref{fig:diag-4}c) is also unimportant, since it 
gives only $S \simeq (N/\pi)(g^2/16\pi^2)$, which is always much smaller 
than the leading contribution of Eq.~(\ref{eq:S-estimate}) (it is simply 
a higher order effect with the propagator of an internal elementary 
gauge boson corrected by the dynamics of $G$).  This diagram does not 
give contributions to the $T$ or $U$ parameters because generating them 
requires at least one additional insertion of the gray disk, to pick 
up the effect of custodial breaking encoded in the absence of elementary 
charged $SU(2)_R$ gauge bosons in the spectrum.

The second diagram (Fig.~\ref{fig:diag-4}b) represents a correction 
coming entirely from the $G$ sector.  Thus, as in the tree-level case, 
it does not give contributions to $T$ or $U$ parameter.  Let us now 
focus on the contribution to the $S$ parameter from this diagram. 
In a perturbative expansion, the disk with a hole is the sum of certain 
graphs as shown in Fig.~\ref{fig:diag-5}.
\begin{figure}[t]
\begin{center} 
\begin{picture}(200,135)(115,80)
  \GCirc(100,190){25}{0.8} \GCirc(100,190){9}{1} 
  \Text(150,190)[]{$=$}
  \Gluon(200,199)(200,215){2.6}{2} \Gluon(200,181)(200,165){2.6}{2}
  \CArc(200,190)(25,0,360) \Text(201,200)[lb]{$A^G_\mu$} 
  \CArc(200,190)(9,0,360) \Text(182,211)[rb]{$\psi_G$}
  \Text(240,190)[]{$+$}
  \Gluon(280,199)(280,215){2.6}{2} \Text(281,200)[lb]{$A^G_\mu$} 
  \Gluon(288,186)(302,178){2.6}{2} \Gluon(272,186)(258,178){2.6}{2}
  \CArc(280,190)(25,0,360) \Text(262,211)[rb]{$\psi_G$} 
  \CArc(280,190)(9,0,360)
  \Text(320,190)[]{$+$} \Text(335,190)[l]{$\cdots$}
  \Text(150,135)[]{$\simeq$} 
  \Text(200,135)[]{\Large $\frac{\tilde{g}^4 N^2}{(16\pi^2)^3}$}
  \Text(240,135)[]{$+$} 
  \Text(280,135)[]{\Large $\frac{\tilde{g}^6 N^3}{(16\pi^2)^4}$}
  \Text(320,135)[]{$+$} \Text(335,135)[l]{$\cdots$}
  \Text(150,95)[]{$=$}
  \Text(175,95)[l]{\Large 
    $\frac{1}{16\pi^2}f'(\frac{\tilde{g}^2 N}{16\pi^2})$}
  \Text(280,95)[]{$\sim$} \Text(305,95)[l]{\Large $\frac{1}{16\pi^2}$}
\end{picture}
\caption{The diagrams represented by the gray annulus; $\psi_G$ 
 and $A^G_\mu$ are matter and gauge fields of the $G$ sector.  
 The contribution from this set of diagrams is of order $1/16\pi^2$.} 
\label{fig:diag-5}
\end{center}
\end{figure}
A similar reasoning as before implies that the contribution from this 
annulus is of order $1/16\pi^2$, so that it gives a contribution to 
the $S$ parameter of order $S \simeq 1/\pi$.  It is then clear that, as 
long as $N \simgt 1$, the contribution from Fig.~\ref{fig:diag-4}b is at 
most comparable to the leading contribution of Eq.~(\ref{eq:S-estimate}). 
Note that the condition $N \simgt 1$ is equivalent to the condition that 
the 5D gauge couplings are smaller than the value determined by NDA, 
$(g_L^2+g_R^2) \simlt 16\pi^3/M_*$, and the AdS curvature scale is 
smaller than the cutoff scale of the theory, $\pi k \simlt M_*$ (see 
Eq.~(\ref{eq:def-N})), which we abide by throughout the paper.  The 
precise value of $S$ coming from the diagram of Fig.~\ref{fig:diag-4}b 
depends on how we define the $S$ parameter.  In particular, if we 
define our $S$ parameter to be the deviation from the standard model 
value, this contribution depends on the reference value for the 
physical Higgs-boson mass, $m_H$, arbitrarily chosen to calculate the 
standard model contribution.  However, this dependence on $m_H$ is not 
very important for later arguments in the paper.  This is because, 
unless some calculable negative contribution to $S$ cancels the 
leading contribution of Eq.~(\ref{eq:contr-s}), the contribution 
from Fig.~\ref{fig:diag-4}b becomes important only in the region 
$N \simeq 1$, where all higher order corrections also become 
non-negligible and we are only able to say that the contribution 
to $S$ is of order $1/\pi$. 

The dominant contributions to the $T$ and $U$ parameters come from 
the diagram of Fig.~\ref{fig:diag-4}d, whose size is estimated as 
$(N/16\pi^2)(g^2/16\pi^2)$, where $g \simeq 1$ represents generic 
electroweak gauge couplings.  Since the natural scale of the $G$ sector 
is $m_{\rho}$, this gives contributions to the quantities $\Pi_{XX}(0)$ 
defined by ${\cal L} = -(1/2)(v^2/2 + \Pi_{XX}(0)) A^X_\mu A^X_\mu$ as 
$\Pi_{XX}(0) \simeq (N/16\pi^2)(g^2/16\pi^2)m_{\rho}^2$.  Then, 
according to the definition of $T$, we have 
\begin{equation}
  T \simeq \frac{16\pi}{g^2 v^2} \Pi_{XX}(0)
    \simeq \frac{1}{\pi},
\label{eq:T-estimate}
\end{equation}
where we have used Eq.~(\ref{eq:fpi-mrho}) with $f_\pi = v$.  Again, 
the precise value of $T$, defined as the deviation from the standard 
model prediction, depends on the reference value for $m_H$ used to 
compute the standard model contribution.  For the $U$ parameter, 
we note that the diagram of Fig.~\ref{fig:diag-4}d gives a contribution 
to $\Pi'_{XX}$ of order $(N/16\pi^2)(g^2/16\pi^2)$.  Then, from the 
definition of $U$, we obtain
\begin{equation}
  U \simeq 16\pi\, \Pi'_{XX}
    \simeq \frac{g^2 v^2}{m_1^2}T.
\label{eq:U-estimate}
\end{equation}
This shows that as long as $(gv/m_1)^2 \ll 1$, as in our case, the 
contribution to the $U$ parameter is negligible. 

By deforming the diagrams of Fig.~\ref{fig:diag-4} as in the way we 
deformed Fig.~\ref{fig:diag-1} to Fig.~\ref{fig:diag-3}, we can easily 
see that these diagrams actually correspond in the 5D picture to the 
one-loop diagrams in which the KK gauge bosons circulate in the loop. 
The reason why these diagrams can give only subdominant contributions 
to $S$ is then clear because the size of the loop diagrams is always 
smaller than that of the tree-level effects, unless the quantity 
is first generated at the loop level as in the case of the $T$ and $U$ 
parameters.  (Remember that $N \simgt 1$ corresponds to the condition 
that the 5D gauge couplings, $g_{\rm 5D}$, are smaller than the value 
given by NDA: $g_{\rm 5D}^2 \simlt 16\pi^3/M_*$.)  From the above 
4D analysis, we know that the 5D loop contributions to $S$ and $T$, 
given by $S \simeq T \simeq 1/\pi$, do not depend on $N$, i.e. the mass 
scale of the KK excitations.  In fact, in the 5D theory we find that 
the 3-point couplings involving a lowest-mode (our $W$ and $Z$) and two 
KK gauge bosons scale as $1/\sqrt{N} \propto m_1/v$.  Thus they cancel 
the mass suppression arising from the KK gauge-boson propagators when 
we calculate $S$ and $T$, making these contributions non-decoupling.

We close this subsection with a final important remark.  In the above 
discussion, we have presented our analysis assuming that matter fields 
in the $G$ sector, $\psi_G$, transform as the fundamental representation 
under $G$.  We have also implicitly assumed that the number of these 
fields are of $O(1)$ and not of $O(N)$.  These assumptions, however, 
will most likely be violated in the present theory because it must 
be nearly conformal above the scale $\Lambda \approx k'$, which 
requires large representations or a large number of matter fields 
to make the beta function nearly vanishing.  Nevertheless, this will 
not change any of our physical conclusions described here and below. 
Let us, for example, consider the case where the matter sector 
consists of $O(1)$ number of $\psi_G$'s that transform as the 
adjoint representation under $G$.  In this case the gray disks in 
Figs.~\ref{fig:diag-1}~--~\ref{fig:diag-3},~\ref{fig:diag-4}c~and~%
\ref{fig:diag-4}d are replaced by gray spheres, which give 
contributions of order $N^2/16\pi^2$, and consequently $N$ in 
Eqs.~(\ref{eq:S-estimate},~\ref{eq:fpi-mrho},~\ref{eq:def-N}) 
must be replaced by $N^2$.  Similarly, the gray annuli in 
Figs.~\ref{fig:diag-4}b~and~\ref{fig:diag-5} are replaced by gray 
tori.  However, these replacements do not change any of the relations 
between the physical quantities --- it simply says that the quantity 
called $N$ before must actually be identified as $N^2$. (Note that 
{\it not} all of the $N$'s must be replaced by $N^2$. For example, 
the expansion in Figs.~\ref{fig:diag-2}~and~\ref{fig:diag-5} are 
given by $(N^2/16\pi^2)f(\tilde{g}^2N/16\pi^2)$ and $(1/16\pi^2)
f'(\tilde{g}^2N/16\pi^2)$, respectively, and not $(N^2/16\pi^2)
f(\tilde{g}^2N^2/16\pi^2)$ and $(1/16\pi^2)f'(\tilde{g}^2N^2/16\pi^2)$. 
Basically, $N$ appearing in $\L \equiv \tilde{g}^2N/16\pi^2$ is not 
replaced by $N^2$.)  One consequence of considering adjoint matter is that 
the corresponding gravitational theory now seems to be closed string theory, 
because the sphere and torus do not have any edge identified as the endpoint 
of strings (for another implication, on an understanding of the fermionic 
KK towers, see footnote~\ref{ft:top-kk}).  The situation is similar in 
the case of $O(N)$ number of $\psi_G$'s transforming as the fundamental 
representation: the quantity called $N$ should be identified as $N^2$, 
although the disks and annuli in this case are not replaced by other 
objects and the corresponding gravitational theory is still an open 
string theory.  In the rest of the paper, we keep using $N$ as it 
appeared in the heuristic presentation in this subsection, as it will 
not change any of the physical results.  The reader who wants a more 
precise picture, however, should understand it appropriately as the 
square of the number of ``colors'' of the gauge group $G$.

\subsection{Comparison with experimental data: $S$ parameter}

Current experimental data already give strong constraints on possible 
new physics at the TeV scale.  For example, the absence of FCNCs other 
than those arising from the standard model strongly constrains the flavor 
structure for the TeV physics.  Here we concentrate on the constraints 
arising from the precision electroweak data, especially those on $S$ and 
$T$ oblique parameters.  The issue of flavor changing processes will be 
discussed in the next section.

We first note that our $S$ and $T$ parameters are defined as the 
deviations from the standard model values.  The standard model 
contributions to the vacuum polarizations are calculated once the 
mass of the physical Higgs boson, $m_H$, is specified.  On the 
other hand, in our theory there is no Higgs boson in the spectrum 
so that the contributions to the vacuum polarizations do not depend 
on such parameter.  This means that $S$ and $T$, defined as the 
differences between the vacuum polarizations in our theory and those 
in the standard model, depend on $m_H$, which is arbitrary chosen 
to calculate the standard model contribution.  Specifically, $S$ 
and $T$ arising from the diagrams of Figs.~\ref{fig:diag-4}b~and~%
\ref{fig:diag-4}d depend on the parameter $m_H$.  Of course, this 
dependence on $m_H$ is not physical --- the experimental constraints 
on $S$ and $T$ also depend on $m_H$, and the physical constraints 
on (the $G$ sector of) our model do not depend on the arbitrary 
parameter $m_H$.  Treating this issue correctly would become important 
when we aim to make a precise comparison between the predictions 
of the theory and experiments. However, we do not need such a 
precision for the purpose here, as we do not attempt to make the 
comparison between the theory and experiment at the level of 5D 
one-loop contributions.  Rather, we discuss general implications 
of the results in the previous subsection, focusing on the large 
leading-order contribution. 

We therefore regard that the theory is successful if it gives 
sufficiently small values of $S$ and $T$.  Specifically, we here 
take a somewhat conservative criterion $S, T \simlt 1/\pi$, and 
consider whether the contribution from the gauge sector derived in 
the previous subsection satisfies it.  Below, we will explicitly see 
that the minimal theory with a perturbative 5D energy region fails 
to pass this test, which implies that either the extension of the 
model or the deviation from the perturbative 5D picture is necessary 
for the theory to be viable.  As we will see in the next section, 
our conclusion is not changed by including contributions from the 
matter sector.

With our weak criterion $S, T \simlt 1/\pi$, only the dangerous 
contribution is the leading-order contribution to the $S$ parameter 
given in Eq.~(\ref{eq:S-estimate}) or Eq.~(\ref{eq:contr-s}). 
These equations imply that, if the coefficient $c_S$ is order 1, 
we need to go to the parameter region $N \simeq 1$, which requires 
that the 5D theory is strongly coupled already at the scale of the 
lowest excitation.  To see the situation more quantitatively, however, 
we have to calculate the coefficient $c_S$.  This can be done in 5D 
by solving the equations of motion for the gauge fields at tree level.

In order to analyze the most general situation, we add the following 
gauge kinetic terms localized on the TeV brane
\begin{equation}
  {\cal S} = \int\!d^4x \int\!dy \sqrt{-g_{\rm ind}} 
    \delta(y-\pi R) \Biggl[ \sum_{a=1}^{3} \Biggl\{
    -\frac{Z_L}{4} F^{La}_{\mu\nu} F^{La}_{\mu\nu}
    -\frac{Z_R}{4} F^{Ra}_{\mu\nu} F^{Ra}_{\mu\nu}
    -\frac{Z_M}{2} F^{La}_{\mu\nu} F^{Ra}_{\mu\nu} \Biggr\}
    -\frac{Z_X}{4} F^X_{\mu\nu} F^X_{\mu\nu} \Biggr],
\label{eq:TeV-gauge-kinetic}
\end{equation}
in addition to the bulk and Planck-brane localized gauge kinetic 
terms, Eq.~(\ref{eq:gauge-kinetic}).  We can now compute the coefficient 
$c_S$ as a function of the parameters of the theory, $1/g_L^2$, $1/g_R^2$, 
$1/g_X^2$, $1/\tilde{g}_L^2$, $1/\tilde{g}_Y^2$, $Z_L$, $Z_R$, $Z_M$, 
and $Z_X$.  The detailed calculation is given in Appendix~A, and 
the result can be summarized as 
\begin{equation}
  S = c_S \frac{N}{\pi},
\label{eq:final-S}
\end{equation}
where 
\begin{equation}
  N = \frac{16\pi^2}{(g_L^2+g_R^2)k},
\qquad
  c_S = \frac{3}{4} 
    + \frac{g_L^2k \cdot g_R^2k}{(g_L^2+g_R^2)k}(Z_L+Z_R+Z_M).
\label{eq:final-N-cS}
\end{equation}
In the absence of the TeV-brane operators, $Z_L=Z_R=Z_M=0$, we find 
that $c_S = 3/4$ and $N$ must actually be small, $N \simeq 1$, 
for the theory to be viable.  The smallest value of $N$ is obtained 
at the largest values for the bulk gauge couplings.  Suppose that 
one of $g_L$ or $g_R$ becomes strongly coupled at the cutoff scale 
of the 5D theory, $M_*$.  This implies that at least one of $g_L$ 
or $g_R$ is as large as $g_{L,R}^2 \simeq 16\pi^3/M_*$, and thus 
we obtain $N \simeq M_*/(\pi k)$.  Therefore, we find that the 
contribution to $S$ becomes sufficiently small only when the 5D 
cutoff scale is lowered down to the scale of AdS curvature (i.e. 
the IR cutoff scale, $M'_*$, is lowered to the scale close to the 
mass of the first KK resonance, $m_1 \simeq \pi k'$).

At first sight, the above conclusion seems to change if we introduce 
the TeV-brane gauge kinetic terms, $Z_L, Z_R, Z_M \neq 0$, because of 
the second term in the equation for $c_S$ in Eq.~(\ref{eq:final-N-cS}). 
A careful study, however, shows that the conclusion actually persists 
even in the presence of the TeV-brane terms.  To see this, we first 
rewrite $Z_L, Z_R$ and $Z_M$ as $Z_L \equiv \delta_L/16\pi^2$, 
$Z_R \equiv \delta_R/16\pi^2$ and $Z_M \equiv \delta_M/16\pi^2$, 
respectively.  The NDA values for these coefficients are then 
represented as $\delta_L, \delta_R, \delta_M = O(1)$.  An important 
point here is that these parameters cannot take large negative values, 
because it would lead to a ghost below the IR cutoff scale, $M_*'$. 
We thus have constraints $\delta_L, \delta_R, \delta_M \simgt -1$ from 
the consistency of the theory, which gives a strong restriction on the 
possibility that the second term in the expression of $c_S$ cancels 
the first term and gives smaller values of $c_S$ (and thus allows larger 
values of $N$).  The second term of $c_S$ in Eq.~(\ref{eq:final-N-cS}) 
becomes most negative when at least one of $g_L$ or $g_R$ takes the 
largest value, in which case 
\begin{equation}
  c_S \simeq \frac{3}{4} + \frac{\pi k}{M_*}(\delta_L+\delta_R+\delta_M).
\label{eq:TeV-corr}
\end{equation}
We thus find that $c_S$ can be much smaller than $O(1)$ only when the 
two scales $M_*$ and $\pi k$ are close.  In fact, Eq.~(\ref{eq:TeV-corr}) 
suggests that the effect encoded in the TeV-brane kinetic terms should 
be regarded as ``stringy corrections'', i.e. the higher order effect in 
the $1/\L$ expansion.

The argument described above explicitly shows that the minimal theory 
with a large perturbative 5D energy interval, $M_* \gg \pi k$, fails 
to comply with precision electroweak data, because then we have 
$c_S \simgt 1$ and $N \gg 1$.  There are then two possibilities to make 
the theory viable.  The first one is to extend the minimal model to 
include a new sector that gives a negative contribution to $S$ and 
cancels the leading gauge contribution of Eq.~(\ref{eq:final-S}).  Such 
a contribution may (effectively) arise from additional matter fields 
(localized to the TeV brane)~\cite{Georgi:1991ci}, additional gauge 
bosons~\cite{Holdom:1990xp} or, perhaps, even from the physics associated 
with the radion field~\cite{Csaki:2000zn} in which case the extension 
of the model may not actually be needed.  In this case, the 5D theory 
can be perturbative up to the cutoff scale $M_*$ which is parametrically 
higher than the AdS curvature scale $\pi k$.  This means that the 
strongly-coupled $G$ sector that breaks electroweak symmetry has a weak 
coupling description over a certain energy interval above the mass of 
the lowest excitation, $\pi k'$, up to some higher energy scale, $M_*'$. 
This type of theory would then allow precise computations of electroweak 
corrections, for example $S$, $T$ and $U$ parameters generated at the 
5D loop level (although for $S$ there are intrinsic uncertainties of the 
same order arising at tree level from operators on the TeV brane).  The 
unitarity of the longitudinal $WW$ scattering amplitudes is recovered 
by the presence of the electroweak KK gauge bosons, instead of the Higgs 
boson~\cite{Csaki:2003dt,Hall:2001tn}. The required cancellation to 
attain these is of order $1/N \simeq \pi k/M_*$.  Thus, in order to 
have a reasonable energy interval where the theory has a weak coupling 
description, say $M_*/(\pi k) \simeq 5$, we only have to invoke the 
cancellation of order $20\%$.

The second possibility is to give up a weak coupling description of 
the theory; specifically, we take $M_* \simeq \pi k$.  In this case 
the mass of the first excited mode, $m_1 \simeq \pi k'$, is close to 
the scale where the theory becomes truly strongly coupled, $M_*'$. 
There is no energy interval where the theory admits a weak coupling 
description, and electroweak radiative corrections cannot be reliably 
computed.  Nevertheless, NDA suggests that in this parameter 
region the corrections to $S$ and $T$ are both of order $1/\pi$. 
Therefore, it does not seem so unnatural that these corrections 
in fact do satisfy the constraints from precision electroweak 
measurements.  The worry, of course, is that because the cutoff scale 
is close to the first KK mass, the 5D field theoretic description of 
the theory may not make much sense.  For instance, if the 5D Planck 
scale $M_5$ is taken close to the 5D cutoff scale, as in the usual 
case, the background AdS solution itself will receive large quantum 
gravitational corrections and our entire treatment will become unreliable. 
However, the 5D Planck scale, $M_5$, may be parametrically larger than 
the cutoff scale, $M_*$.  The argument based on the locality in 5D may 
also persist even for $M_* \simeq \pi k$, as the proper distance for the 
fifth dimension, $\pi R$, is still larger than the cutoff length, $1/M_*$. 
Here we do not try to make further arguments on the viability of this 
parameter region.  A more solid treatment of this region will probably 
require a string theoretic construction of the theory.

Having the above two possibilities in mind, we will discuss further 
phenomenological issues of the theory in the next section.  These 
include flavor violation, top quark phenomenology, and contributions 
of the matter sector to the electroweak oblique parameters.

\section{Fermion Sector and Its Phenomenology}
\label{sec:top}

In this section we discuss the fermion sector of the model described in 
section~\ref{sec:model}.  We focus most of our discussions on the third 
generation quarks since they are most severely constrained by experiments. 
However, some of our analyses, for instance those for mass eigenvalues and 
flavor violation, are also applicable to the first two generation quarks and 
leptons.  A related study on the issue of fermion masses can be found 
in~\cite{Csaki:2003sh}.

\subsection{Basic structure}

Let us start by summarizing the structure of the fermion sector of 
the model given in section~\ref{sec:model}.  A single generation of 
the quark sector consists of the following fermion content:
\begin{eqnarray}
&& q({\bf 2}, {\bf 1}, 1/6) 
     = \pmatrix{ u \cr d }, \qquad
   \psi_{\bar{u}}({\bf 1}, {\bf 2}, -1/6)
     = \pmatrix{ \bar{D} \cr \bar{u} }, \qquad
   \psi_{\bar{d}}({\bf 1}, {\bf 2}, -1/6)
     = \pmatrix{ \bar{d} \cr \bar{U} }, 
\nonumber \\
&& q^c({\bf 2}^*, {\bf 1}, -1/6) 
     = \pmatrix{ u^c \cr d^c }, \qquad
   \psi_{\bar{u}}^c({\bf 1}, {\bf 2}^*, 1/6)
     = \pmatrix{ \bar{D}^c \cr \bar{u}^c }, \qquad
   \psi_{\bar{d}}^c({\bf 1}, {\bf 2}^*, 1/6)
     = \pmatrix{ \bar{d}^c \cr \bar{U}^c }.
\label{eq:quark-rep}
\end{eqnarray}
Here $q$, $\psi_{\bar{u}}$, $\psi_{\bar{d}}$, $q^c$, $\psi_{\bar{u}}^c$ 
and $\psi_{\bar{d}}^c$ represent (left-handed) Weyl fermions; $q$ and $q^c$ 
form a single 5D (Dirac) fermion, and the same is true for $\psi_{\bar{u}}$ 
and $\psi_{\bar{u}}^c$, and for $\psi_{\bar{d}}$ and $\psi_{\bar{d}}^c$.
The numbers in parentheses on the left-hand side of the equations 
represent the quantum numbers under $SU(2)_L \times SU(2)_R \times 
U(1)_X$.  

Under the orbifold boundary conditions the zero modes for the conjugate 
fields are projected out, so that the fields in the lower line of 
Eq.~(\ref{eq:quark-rep}) do not have zero modes.  The zero modes 
for $\bar{U}$ and $\bar{D}$ also get masses by marrying with the 
Planck-brane fields $\psi'_{\bar{u}}$ and $\psi'_{\bar{d}}$ (see 
discussions below Eq.~(\ref{eq:fermion-rep})).  Therefore, before 
turning on the effect of electroweak symmetry breaking, the three 
Weyl-fermion fields $q$, $\bar{u}$ and $\bar{d}$ are massless. 
These fields have the quantum numbers $({\bf 2},1/6)$, $({\bf 1},-2/3)$ 
and $({\bf 1},1/3)$ under $SU(2)_L \times U(1)_Y$ and are identified 
as the quarks in the standard model.  The fermion KK towers consist 
of all the fields listed in Eq.~(\ref{eq:quark-rep}).

Once the effect of electroweak symmetry breaking is introduced through the 
operators in Eq.~(\ref{eq:yukawa}), the standard-model quarks receive masses, 
which depend on the bulk mass parameters $c$ for $q$, $\psi_{\bar{u}}$, 
$\psi_{\bar{d}}$ as well as the TeV-brane couplings, $y_u$ and $y_d$. 
These masses, together with the KK tower mass spectrum, are worked out 
in Appendix~B. They are determined by the condition
\begin{eqnarray}
  && \Biggl( J_{c_L-\frac{1}{2}}\Bigl(\frac{m}{k'}\Bigl) 
    - \frac{J_{c_L-\frac{1}{2}}\bigl(\frac{m}{k}\bigl)}
      {Y_{c_L-\frac{1}{2}}\bigl(\frac{m}{k}\bigl)}
      Y_{c_L-\frac{1}{2}}\Bigl(\frac{m}{k'}\Bigl) \Biggr)
  \Biggl( J_{c_R-\frac{1}{2}}\Bigl(\frac{m}{k'}\Bigl) 
    - \frac{J_{c_R-\frac{1}{2}}\bigl(\frac{m}{k}\bigl)}
      {Y_{c_R-\frac{1}{2}}\bigl(\frac{m}{k}\bigl)}
      Y_{c_R-\frac{1}{2}}\Bigl(\frac{m}{k'}\Bigl) \Biggr)
\nonumber\\
  && - |\lambda|^2 
  \Biggl( J_{c_L+\frac{1}{2}}\Bigl(\frac{m}{k'}\Bigl) 
    - \frac{J_{c_L-\frac{1}{2}}\bigl(\frac{m}{k}\bigl)}
      {Y_{c_L-\frac{1}{2}}\bigl(\frac{m}{k}\bigl)}
      Y_{c_L+\frac{1}{2}}\Bigl(\frac{m}{k'}\Bigl) \Biggr)
  \Biggl( J_{c_R+\frac{1}{2}}\Bigl(\frac{m}{k'}\Bigl) 
    - \frac{J_{c_R-\frac{1}{2}}\bigl(\frac{m}{k}\bigl)}
      {Y_{c_R-\frac{1}{2}}\bigl(\frac{m}{k}\bigl)}
      Y_{c_R+\frac{1}{2}}\Bigl(\frac{m}{k'}\Bigl) \Biggr) = 0,
\label{eq:det-mass}
\end{eqnarray}
where $m$ represents the mass eigenvalues: the masses for our quarks 
and the KK towers are given as the solutions to this equation. 
Here $J_\nu(x)$ and $Y_\nu(x)$ are Bessel functions, $c_L$ and $c_R$ 
are the bulk masses for the left-handed and right-handed fermions, 
and $\lambda$ represents the size of the TeV-brane operators.  For example, 
if we want to know the mass eigenvalues of the up-type quark KK tower, we 
substitute $c_L = c_q$, $c_R = c_{\psi_{\bar{u}}}$ and $\lambda = y_u v_H$ 
in Eq.~(\ref{eq:det-mass}) and solve. In the case of the down-type quark, 
we use $c_L = c_q$, $c_R = c_{\psi_{\bar{d}}}$ and $\lambda = y_d v_H$. 
For the leptons, $c_L = c_l$, $c_R = c_{\psi_{\bar{e}}}$ and 
$\lambda = y_e v_H$.

For the first two generation quarks (and leptons), we choose $c_L$ and 
$c_R$ sufficiently larger than $1/2$.  In this case the lightest mass 
eigenvalues, i.e. the masses of our quarks and leptons, are suppressed 
by a large factor, $m \sim k'(k'/k)^{c_L+c_R-1}$, explaining the 
hierarchy among quark and lepton masses~\cite{Gherghetta:2000qt}. 
Because we do not want to have such a suppression for the third 
generation quarks, especially for the top quark, we choose $c_q$ and 
$c_{\psi_{\bar{u}}}$ for the third generation to be smaller than $1/2$. 
The third generation $c_{\psi_{\bar{d}}}$ is taken to be larger than 
but close to $1/2$ so that the bottom quark mass is not too suppressed. 
More detailed discussions on phenomenology of the third generation sector, 
including the mass spectrum of the top-quark KK tower, appear in the 
following subsections. 

\subsection{Constraints from flavor violation}
\label{subsec:fv}

\begin{figure}[t]
\begin{center}
\begin{picture}(270,110)(145,117)
  \Text(190,137)[t]{\large (a)}
  \Photon(205,190)(235,190){3}{3.75} 
  \Text(235,198)[b]{\small $W_\mu, B_\mu$} 
  \Line(190,190)(168,229)  \Text(165,229)[r]{$\psi$}
  \Line(190,190)(168,151)  \Text(165,151)[r]{$\psi$}
  \GCirc(190,190){15}{0.8} 
  \Text(375,137)[t]{\large (b)}
  \Line(363,202)(338,229)  \Text(336,229)[r]{$\psi$}
  \Line(363,178)(338,151)  \Text(336,151)[r]{$\psi$}
  \Line(387,202)(412,229)  \Text(416,229)[l]{$\psi$}
  \Line(387,178)(412,151)  \Text(416,151)[l]{$\psi$}
  \GCirc(375,190){18}{0.8}
\end{picture}
\caption{Diagrams giving flavor non-universal gauge interactions~(a) 
 and four-fermion operators~(b).  The gray disks represent the 
 dynamics of the $G$ sector.}
\label{fig:diag-fnu}
\end{center}
\end{figure}
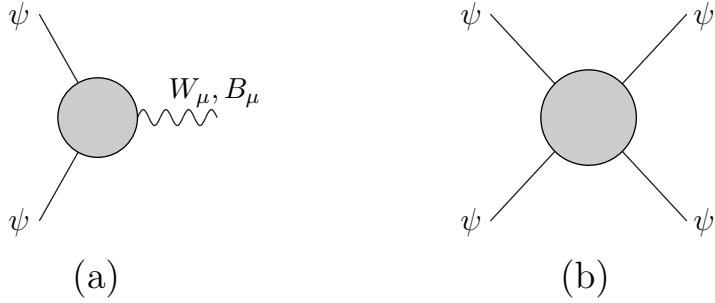

Since the electroweak gauge symmetry is broken by boundary conditions 
or a large VEV of the brane-localized field $H$, the wavefunctions 
of the $W$ and $Z$ bosons in our theory are not flat in the extra 
dimension.  This generically introduces non-universality of the 
electroweak gauge couplings depending on the bulk fermion mass 
parameters, because the 4D gauge couplings are obtained by convolving 
the wavefunctions of the corresponding fermion with the gauge boson, 
which are not universal for fermions having different values of the 
bulk masses.  To estimate the size of this non-universality, we 
first consider the theory in the 4D picture.  In this picture the 
non-universality for the electroweak gauge couplings is caused by 
the diagram as shown in Fig.~\ref{fig:diag-fnu}a.  Here, the 
solid external lines are fermion fields and the wavy external 
line is the electroweak gauge boson.  The gray disk at the center 
represents the dynamics of the $G$ sector. 

What is the coupling of the external fermion lines (elementary fermion 
fields) to the $G$ sector?  In the 4D picture, an elementary fermion 
field $\psi$ couples to the $G$ sector through the interaction like 
${\cal L}_{\rm 4D} \sim \psi {\cal O}_\psi$, where ${\cal O}_\psi$ is 
an operator which consists of fields in the $G$ sector and has the quantum 
numbers of $\psi$ conjugate.  The dimension of this operator is related to 
the bulk mass $c$ of the 5D field corresponding to $\psi$.  For $c \geq -1/2$ 
it is given by $[{\cal O}_\psi] = c+2$~\cite{Henningson:1998cd}. Therefore, 
in terms of the canonically normalized field $\psi$, the coupling of $\psi$ 
and the $G$ sector can be written as 
\begin{equation}
  {\cal L}_{\rm 4D} = \frac{\eta}{k^{c-1/2}} \psi {\cal O}_\psi,
\label{eq:fermion-dual}
\end{equation}
where $\eta$ is a dimensionless coupling constant.  For $c > 1/2$ 
this is an irrelevant operator and the coupling $\eta$ is of order 
one.  For $c < 1/2$, the interaction in Eq.~(\ref{eq:fermion-dual}) 
is relevant, and the coupling $\eta$ runs with energy 
as $\eta(\mu) \sim (k/\mu)^{c-1/2}$, which implies that 
$\eta/k^{c-1/2}$ provides an order-one insertion at any given 
energy~\cite{Contino:2003ve}.  For $c=1/2$ the operator is 
marginal and $\eta$ is given by $\eta \sim 1/(\ln(k/k'))^{1/2}$ 
at the electroweak scale.  For $c$ larger than but close to $1/2$, 
$\eta \simeq ((2c-1)/(1-(k'/k)^{2c-1}))^{1/2}$.

We can now estimate the non-universality arising from the diagram 
of Fig.~\ref{fig:diag-fnu}a.  For $c > 1/2$ the fermion field is 
attached to the gray disk with the factor $\eta/k^{c-1/2}$. 
The gray disk contribute as $N/16\pi^2 \sim 1/(g_L^2+g_R^2)k$.%
\footnote{This estimate may be justified by the following argument. 
The fact that the 4D gauge couplings receive contributions from 
the $G$ sector proportional to $N$ implies that the $G$ matter charged 
under $SU(2)_L \times U(1)_Y$ is in fundamental representations of 
$G$.  Then, the fields circulating on the edge of the disk must be 
fundamental representations of $G$ (the edge between the two external 
fermion lines is a scalar), and the standard counting in large $N$ 
for the group-theoretical factor gives $N/16\pi^2$.  An alternative 
possibility is that matter in the $G$ sector, $\psi_G$, is in the 
adjoint representation under $G$ (see the last paragraph of 
section~\ref{subsec:structure-1}).  In this case, the gray disk becomes 
a gray sphere and contributions from the $G$ sector to the 4D gauge 
couplings are proportional to $N^2$.  The $N$ power counting perfectly 
works just by replacing $N$ by $N^2$.  The adjoint matter also provides 
an understanding of the KK towers of the fermion fields as bound states 
of $\psi_G$ and $A^G_\mu$ (the gauge bosons of $G$).  A fundamental 
scalar field is not necessary in this case.  \label{ft:top-kk}} 
Multiplying these factors and supplying the dimension by $k'$ then 
leads to the non-universality of the gauge coupling $\delta g$:
\begin{equation}
  \frac{\delta g}{g} = a \frac{\eta^2 k'^{2c-1}}{(g_L^2+g_R^2)k^{2c}}.
\end{equation}
where $a$ is an $O(1)$ constant.  In fact, this parametric dependence 
can be recovered in the 5D calculation.  Defining the non-universality 
$\delta g$ as the deviation of the gauge coupling from the case 
of Planck-brane localized fermions, we obtain it by convolving the 
wavefunctions of the matter zero mode and the electroweak gauge boson. 
We find that $a \simeq 1$--$10$ that depends quite weakly on $c$ (the 
$a$ here contains the possible effect from brane couplings (see 
Eq.~(\ref{eq:fnu-zbb-1}) below), which arises from the fact that our 
$W$ and $Z$ bosons are mixtures of elementary and composite states).
This constitute a small non-universality, and the constraints from 
flavor violating processes for the first two generation quarks (as well 
as leptons) are evaded relatively easily~\cite{Nomura:2003du}, especially 
when $g_R$ is large, i.e. the mass of the first KK gauge boson becomes 
large. Similarly, non-universal four-fermion operators generated 
by the exchange of the KK gauge bosons, represented by the diagram of 
Fig.~\ref{fig:diag-fnu}b, are also quite small for the first two generations. 
They are given by $\delta G_F \simeq \eta^4 (k'/k)^{4c-2}/((g_L^2+g_R^2)kk'^2)$, 
where $\delta G_F$ represents the coefficients of the flavor violating 
four-fermion operators obtained after integrating out the KK gauge bosons.%
\footnote{Diagrams similar to Figs.~\ref{fig:diag-fnu}a~and~%
\ref{fig:diag-fnu}b but with $\psi$'s replaced by gauge fields 
give corrections to the gauge three-point and four-point 
interactions, $ZWW$, $WWWW$ and $ZZWW$.  These corrections are 
of order $\delta g/g \simeq (N/16\pi^2)g^2 \simeq g^2 v^2/m_1^2$.}

For the third generation quarks, $c_q$ and $c_{\psi_{\bar{u}}}$ must be  
smaller than $1/2$ in order to give a large enough mass to the top quark. 
The value of $c_q$ is then constrained by the flavor violating coupling of 
the left-handed $b$ quark to the $Z$ boson.  Performing a full 5D calculation, 
we find that the relevant flavor violation is parameterized by
\begin{equation}
  \frac{\delta g^b_L}{g^b_L} = \frac{f}{(g_L^2+g_R^2)k} 
    \left( \frac{g_L^2}{\pi R\, g^2} \right),
\label{eq:fnu-zbb-1}
\end{equation}
where $g^b_L$ is the coupling of the left-handed bottom quark to the 
$Z$ boson, and $f = f(c_L,c_R,\lambda)$ is a function of $c_L = c_q$, 
$c_R = c_{\psi_{\bar{d}}}$ and $\lambda = y_d H$.  The last factor 
captures the dependence of $\delta g^b_L/g^b_L$ on the brane 
couplings, becoming $1$ in their absence (see Eq.~(\ref{eq:corresp})). 
For $c_R > 1/2$ and $\lambda \simlt 1$, the coefficient $f$ effectively 
depends only on $c_L$.  The dependence is roughly given by
\begin{equation}
  f(c_L) \approx -a'\eta^2 
    = -a' \frac{1-2c_L}{1-(k'/k)^{1-2c_L}},
\label{eq:fnu-zbb-2}
\end{equation}
where the dependence of $a'$ on $c_L$ is rather weak (for instance, 
$6 \simlt a' \simlt 7$ in the parameter region we are interested, 
$0.3 \simlt c_L \simlt 0.5$).
\begin{figure}[t]
\begin{center}
  \epsfig{file=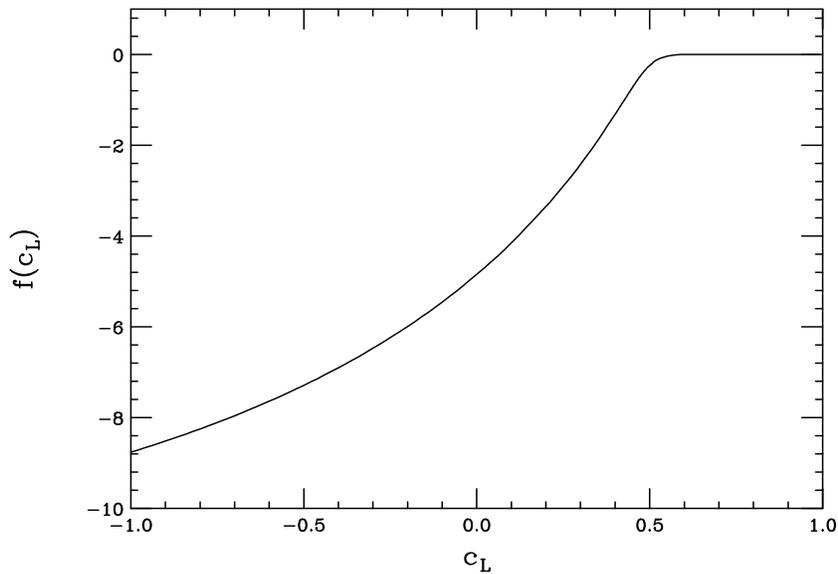,width=7.5cm,height=11cm,angle=90}
\caption{The function $f$ defined in Eq.~(\ref{eq:fnu-zbb-1}), which 
 determines the deviation of $g^b_L$ from its standard model value.}
\label{fig:fcl}
\end{center}
\end{figure}
In Fig.~\ref{fig:fcl} we plot $f$ as a function of $c_L$, calculated in 
the 5D picture.  The $Z \rightarrow b\bar{b}$ decay constrains the quantity 
$\delta g^b_L/g^b_L$ to be less than a percent level.  In our theory, 
$(g_L^2+g_R^2)k \simeq 16\pi^2/N$, and in the case that the brane couplings 
are given by Eqs.~(\ref{eq:brane-kin-TeV}) with $1/\tilde{g}_{L,0}^2 \simeq 
1/\tilde{g}_{Y,0}^2 \simeq 1/16\pi^2$, we find from Eq.~(\ref{eq:fnu-zbb-1}) 
and Fig.~\ref{fig:fcl} that $c_L$ must satisfy $c_L \simgt 0.3, 0.44$ and 
$0.47$ for $N \simeq 1, 3$ and $5$, respectively.  Therefore, to avoid 
having to fine-tune the parameter $c_L$ to be very close to $1/2$, smaller 
values of $N$ are preferred.  This points in the same direction as the 
leading-order $S$ parameter constraint, although the constraint from 
$Z \rightarrow b\bar{b}$ is weaker and allows a reasonable energy interval 
for the weakly coupled 5D description, e.g.~$N \simeq 3$.  In the case 
of strong bulk gauge couplings, Eq.~(\ref{eq:fnu-zbb-1}) gives about 
one percent deviation for $c_L \simeq 0.49$ for $1/g_{L,R,X}^2 \simeq 
M_*/16\pi^3$.  Smaller values of $c_L$, however, are possible if $g_L$ 
takes somewhat smaller values, e.g. $g_L^2 \simlt 8\pi^3/M_*$, or there 
is some cancellation from unknown strong coupling dynamics.

\subsection{Top quark phenomenology}

Obtaining a large enough mass for the top quark is a non-trivial 
issue in any theory with dynamical electroweak symmetry breaking. 
In this subsection we discuss the spectrum of the top quark and 
its KK tower.  We also address the resulting top-quark phenomenology. 

The relevant parameters for the top-quark sector are $c_L = c_q$, 
$c_R = c_{\psi_{\bar{u}}}$ and $\lambda = y_u H$ for the third 
generation.  As seen in the previous subsection, the parameter $c_L$ 
must be close to $1/2$ to avoid the conflict with the observed 
$Z \rightarrow b\bar{b}$ decay rate.  The parameter $c_R$ is less 
constrained and can take much smaller values.  For fixed values of 
$c_L$ and $c_R$, the spectrum of the top KK tower shows the following 
behavior as a function of $\lambda$.  For $\lambda = 0$, the spectrum 
consists of two decoupled towers for the 5D fields $u$ and $\bar{u}$. 
Each tower has a zero mode, which is a Weyl fermion, and a tower of 
Dirac fermions; the towers for $u$ and $\bar{u}$ have identical masses 
for $c_L = c_R$, but in general have slightly different masses for 
$c_L \neq c_R$.  Therefore, the overall spectrum for $\lambda = 0$ 
can be described as follows: there is a Dirac fermion at the massless 
level, which consists of two Weyl-fermion zero modes of $u$ and $\bar{u}$, 
and the KK tower has two Dirac fermions at each level, arising from 
$u$ and $\bar{u}$, whose masses are degenerate for $c_L = c_R$ but not 
in general.  When we turn on $\lambda$ by a small amount ($\lambda \ll 1$), 
the Dirac fermion that is massless for $\lambda = 0$ receives a mass 
proportional to $\lambda$.  Meanwhile, the two Dirac fermions at each 
level become increasingly split: the lightest of the two becomes lighter 
and the heaviest becomes heavier.  For very large $\lambda$ ($\lambda 
\gg 1$), the masses for all the states become constant.  In particular, 
the mass of the formerly massless state approaches to a constant value 
of order $k'$.  The lightest of the first excited states becomes close 
to this state in mass.  They become degenerate at $\lambda \rightarrow 
\infty$ for $c_L = c_R$, but not for $c_L \neq c_R$.

\begin{figure}[t]
\begin{center}
  \epsfig{file=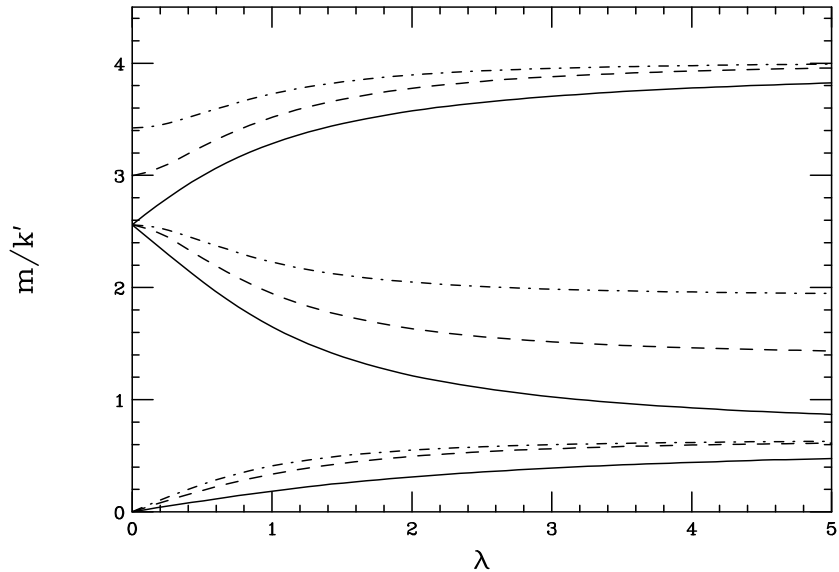,width=7.5cm,height=11cm,angle=90}
\caption{The spectra for the top KK tower.  The horizontal axis 
 is $\lambda = y_u H$, and the vertical axis is the masses in 
 units of $k'$.  The solid, dashed and dot-dashed lines are 
 for $(c_L,c_R) = (0.4,0.4)$, $(0.4,0.1)$ and $(0.4,-0.2)$, 
 respectively.}
\label{fig:topKK-1}
\end{center}
\end{figure}

The behavior of the mass eigenvalues described above is plotted in 
Fig.~\ref{fig:topKK-1} for $c_L = 0.4$ with three different values 
of $c_R$:~$0.4, 0.1, -0.2$. 
In the figure, we have plotted the mass eigenvalues of the top KK 
tower in units of $k'$, obtained by solving Eq.~(\ref{eq:det-mass}) 
for given values of $c_L$ and $c_R$, as a function of $\lambda$.  
The solid, dashed and dot-dashed lines represent the spectra for 
$c_R = 0.4$, $0.1$ and $-0.2$, respectively.  We can see from 
the figure that the masses approach to constant values for 
$\lambda \rightarrow \infty$.  For the lowest mass eigenstate, 
which we identify as the standard-model top quark, the mass 
approaches to $m/k' \simeq 0.65$ for $\lambda \rightarrow \infty$ 
regardless of the value of $c_R$, although the value of $m/k'$ 
obtained for $\lambda = O(1)$ depends on $c_R$.  This is because 
for $\lambda \rightarrow \infty$ the mass eigenvalues are determined 
by the condition that the second term in the left-hand-side of 
Eq.~(\ref{eq:det-mass}) is vanishing; namely, the sum of the solutions 
of $J_{c+1/2}(m/k')-(J_{c-1/2}(m/k)/Y_{c-1/2}(m/k))Y_{c+1/2}(m/k')=0$ 
for $c = c_L$ and $c_R$.  This in turn implies that the maximum value 
of the top-quark mass, i.e. the mass at $\lambda \rightarrow \infty$, 
is determined by the value of the largest of $c_L$ and $c_R$ ($c_L$ 
in the case of Fig.~\ref{fig:topKK-1}), since the value of $m/k'$ 
obtained as the lowest mass solution of the above equation decreases 
for increasing value of $c$.  Specifically, the maximum value of the 
top-quark mass is given by $m/k' \simeq 0.25, 0.45, 0.65, 0.80$ and 
$0.95$ for ${\rm max}\{c_L, c_R\} = 0.5, 0.45, 0.4, 0.35$ and $0.3$, 
respectively.  Therefore, there is a tension between the top-quark 
mass and the constraint from $Z \rightarrow b\bar{b}$ discussed at 
the end of the previous subsection, as $Z \rightarrow b\bar{b}$ 
provides the lower limit for the value of $c_L$~\cite{Agashe:2003zs}. 
Since the value of $k'$ in our theory is $k' \simeq 2\pi v/\sqrt{N}$ 
(from Eqs.~(\ref{eq:corresp})~and~(\ref{eq:final-N-cS})), we find that 
$N$ must satisfy $N \simlt 4$ to obtain a large enough top-quark mass, 
$m_t \simeq 165~{\rm GeV}$ for $\overline{\rm MS}$ without QCD 
radiative corrections.%
\footnote{Introducing Planck-brane localized kinetic terms for 
the left- and right-handed top quarks does not significantly 
modify the argument, unless there are unnatural cancellations.}
We also find from Fig.~\ref{fig:topKK-1} that for sufficiently 
large values for $c_R$ the mass of the first top-quark KK tower is 
significantly lighter than that of the first gauge-boson KK tower, 
$m_1 \simeq 2.4 k'$, in the region $\lambda \gg 1$.  This feature, 
however, is lost for $\lambda \simlt 1$ or smaller values of $c_R$.

\begin{figure}[t]
\begin{center}
  \epsfig{file=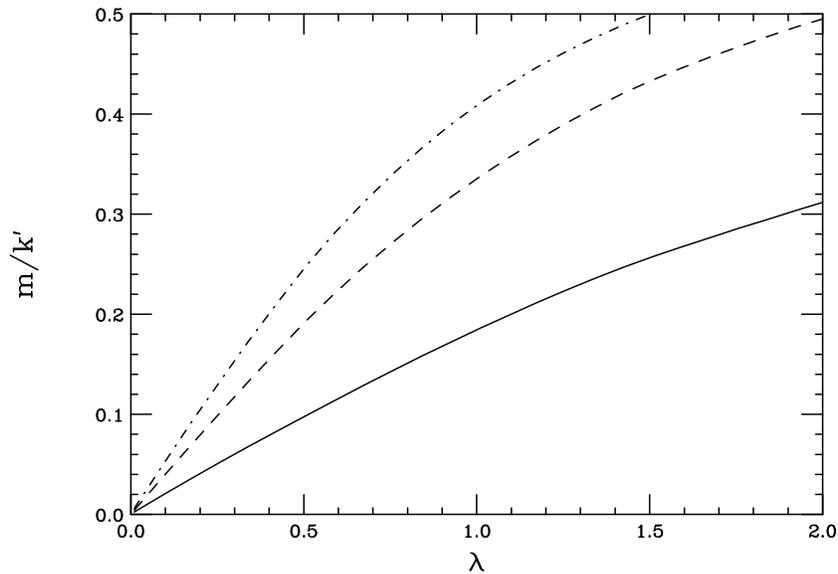,width=7.5cm,height=11cm,angle=90}
\caption{The same as Fig.~\ref{fig:topKK-2} but for $0 \leq 
 \lambda \leq 2$ and $0 \leq m/k' \leq 0.5$.}
\label{fig:topKK-2}
\end{center}
\end{figure}

In Fig.~\ref{fig:topKK-2}, we present a magnification of the lower-left 
corner of Fig.~\ref{fig:topKK-1}. 
\ From this figure we can obtain $\lambda$ for a given value of $N$. 
Let us, for example, consider the two cases of $N \simeq 1$ and $3$. 
In these cases, the values of $k'$ are given by $k' \simeq 1100~{\rm GeV}$ 
and $630~{\rm GeV}$, respectively, requiring $m/k' \simeq 0.15$ and 
$0.26$ to reproduce the observed top-quark mass.  These values are 
easily obtained for $c_R = \{ -0.2, 0.1, 0.4 \}$ by choosing $\lambda 
\simeq \{ 0.3, 0.4, 0.8 \}$ and $\{ 0.5, 0.7, 1.5 \}$, respectively. 
It is interesting that these values of $\lambda$ are what one would 
naively expect from dimensional grounds: $y_u \approx 1/M_*$ and 
$H \approx M_*$.  This is also true for larger values of $c_L$ and for 
the case of strong bulk gauge couplings. For example, for $c_L = 0.49$ 
and $N \simeq 1$ one only needs $\lambda \simeq \{ 0.7, 0.9, 1.9 \}$ 
for $c_R = \{ -0.2, 0.1, 0.4 \}$ to obtain the observed top-quark mass.

We finally discuss phenomenological issues in the top sector.  As 
the $W$ and $Z$ wavefunctions are not flat in the extra dimension, 
the couplings of these gauge bosons to the top quark deviate 
from their standard model values.  For the left-handed top quark, 
the deviation of the coupling to $Z$ is roughly given by 
Eq.~(\ref{eq:fnu-zbb-1}) and at a level of a few percent.  The 
deviation of the $t_L b_L W$ coupling from the standard model is 
also at this level.  Therefore, these effects are not constrained 
by the present experimental data.  The effects on the right-handed 
top quark can be much larger because it has smaller values of $c$.  
For example, smaller values of $c$ induce the interaction of the 
right-handed top quark $t_R$ to the $W$ boson.  However, this 
interaction couples $t_R$ and $W$ only with the heavy field $\bar{B}$ 
and its KK tower, denoted as $\bar{D}$ in Eq.~(\ref{eq:quark-rep}), 
so it is irrelevant at low energies.  A particularly interesting effect 
appears in the right-handed top quark coupling to the $Z$ boson, 
$t_R t_R Z$.  This coupling can have an order-one deviation from the 
standard model value.  The deviation arises mainly from the fact that 
the $Z$ wavefunction inside $A^{R3}_\mu$ has a non-trivial profile, and 
is approximately given by $\delta g^t_R/g^t_R = f'' g_R^2/(g_L^2+g_R^2)$, 
where $f'' \approx (1-2 c_R)/(1-(k'/k)^{1-2c_R})$ with $c_R$ representing 
the bulk mass for $t_R$.  This is in contrast to the case of the 
$b_L b_L Z$ coupling, which arises from the variation of the $Z$ 
wavefunction in $A^{L3}_\mu$ and thus is given by $\delta g^b_L/g^b_L 
= f' g_L^2/(g_L^2+g_R^2)$ with $f' \approx (1-2 c_L)/(1-(k'/k)^{1-2c_L})$ 
(see Eqs.~(\ref{eq:fnu-zbb-1},~\ref{eq:fnu-zbb-2})).  Since $g_R$ 
can generically be larger then $g_L$ (and $c_R$ smaller than $c_L$), 
$\delta g^t_R/g^t_R$ can be large.  It will be interesting to explore 
this coupling in a future $e^+e^-$ linear collider~\cite{Abe:2001nn}.

\subsection{Corrections to oblique parameters}

\ From the analysis in section~\ref{sec:structure} of the electroweak oblique 
corrections generated by the gauge sector, we have found that there are two 
possibilities for making the theory viable.  One is to have $N \simeq 1$, 
i.e. $(g_L^2+g_R^2)k \simeq 16\pi^2$, and the other is to extend the theory 
to give an additional negative contribution to $S$.  In the former case 
the contributions to $S$ and $T$ from the top sector cannot be reliably 
calculated, as the IR cutoff of the theory is close to the scale of 
the first resonance so that there is no energy region where the theory 
has a weak coupling description.  However, an argument similar to the 
gauge case suggests that these contributions are of order $S \approx 
T \approx 1/\pi$.  The relevant diagrams are similar to those in 
Fig.~\ref{fig:diag-4} but with appropriate modifications (internal 
lines should be the top quark or its $SU(2)_R$ partner, there must be 
more insertions of the $G$ dynamics with the chirality flipping effects, 
and so on).  Because the diagrams have insertions of the fermion masses, 
the contributions from the other fermions are much smaller.  Therefore, 
although we do not know the precise contribution from the matter sector, 
we can expect that the theory is still viable in the sense that the 
contributions to the oblique parameters are of order $S,T \simlt 1/\pi$.

In the case with additional negative $S$ contributions, we can have a 
moderate energy interval where the theory is weakly coupled.  In this 
case, the top contribution to $S$ does not give any additional constraint, 
because we have already invoked the cancellation between the gauge and 
the additional negative contributions --- we simply have to make the sum 
of the gauge, top and additional contributions to be smaller than the 
experimental constraint.  As for the top contribution to $T$, it is 
a calculable quantity dominated by the IR region $\approx k'$, due to 
the custodial $SU(2)$ symmetry of the $G$ sector~\cite{Agashe:2003zs}. 
The contribution is roughly given by $T \approx a_t (m_t/k')^2$, where 
$a_t$ is a constant; the value of $a_t$ depends on $c_L$ and $c_R$, and 
may have an enhancement coming from the fact that the $G$ sector (KK 
towers) feels stronger chiral symmetry breaking than the elementary 
sector.  The acceptable values of $T$ from the top sector depend on 
the size of the contributions from the gauge sector and an additional 
sector needed to make $S$ sufficiently small.  However, based on the naive 
estimate, it does not seem implausible to expect that the top contribution, 
together with all the other contributions, actually fit to the data in 
some explicit models.

\subsection{Implications for flavor physics}

As mentioned in section~\ref{subsec:fv}, flavor violation in this model 
arises as a consequence of the need to have different bulk masses for 
the third generation quarks in order to obtain a large top quark mass. 
This non-universality leads to tree-level FCNCs.  These have two main 
manifestations.  First, since the wavefunction of the $Z$ is pushed away 
from the IR brane by the boundary conditions (or a large VEV), there will 
be non-negligible tree-level FCNC couplings of $q^T=(t_L~b_L)^T$ and $t_R$ 
with the $Z$, since these must be localized not too far from this brane. 
We define the effective $Zbs$ coupling by 
\begin{equation}
  {\cal L}_{Zbs} = \frac{g^2}{4\pi^2}\,\frac{g}{2 \cos\theta_W}
    \left(Z_{bs}\,\bar{b}_L\gamma^\mu s_L 
    + Z'_{bs}\,\bar{b}_R\gamma^\mu s_R\right) Z_\mu,
\label{lzbs}
\end{equation}
where $Z_{bs}$ and $Z'_{bs}$ encode both the one-loop standard model 
as well as new physics contributions.  Up to a factor of order one, the 
tree-level FCNC vertex induced by the flavor violating coupling discussed 
in section~\ref{subsec:fv} results in 
\begin{equation}
  \delta Z_{bs} \simeq 
    \Bigl(-\frac{1}{2}+\frac{1}{3}\sin^2\theta_W\Bigr) 
    D_L^{bs}\,\frac{8\pi^2}{g^2} \left( \frac{\delta g^b_L}{g^b_L} \right)
    \simeq -8\pi^2 D_L^{bs}\, \frac{\delta g^b_L}{g^b_L},
\label{dzbs}
\end{equation} 
where $\delta g^b_L/g^b_L$ is given in Eq.~(\ref{eq:fnu-zbb-1}) and takes 
a value of order one percent (or somewhat smaller) in a generic parameter 
region. In Eq.~(\ref{dzbs}), $D_L^{bs}$ corresponds to the $b$-$s$ element 
of the left-handed down-quark rotation diagonalizing the quark masses. 
Thus, with the natural assumption $D^{bs}_L\simeq V_{tb}^*V_{ts}$, the 
correction is of the same order as the standard model contribution to 
this vertex, which is~\cite{Atwood:2003tg} $Z_{bs}^{\rm SM} \simeq -0.04$ 
($Z'^{\rm SM}_{bs} \simeq 0$).  This leads to potentially observable effects 
in $b \to s\ell^+\ell^-$ decays, although the current experimental data, 
$|Z_{bs}| \simlt 0.08$~\cite{Atwood:2003tg}, is not greatly constraining.%
\footnote{This possibility has also been mentioned in~\cite{Agashe} 
in the context of the model of~\cite{Agashe:2003zs}. We thank K.~Agashe 
for pointing out this reference to us.}
The effects are larger in the case that the bound on $Z \rightarrow b\bar{b}$ 
is saturated, i.e. $\delta g^b_L/g^b_L \simeq 1\%$, for example in the 
case of strong bulk gauge couplings. The effect of Eq.~(\ref{dzbs}) also 
contributes to hadronic modes, such as $B \to \phi K_s$, although there 
it must compete with the parametrically larger contributions from gluonic 
penguins. 

The second type of tree-level FCNC effects occur in the interactions of 
third generation quarks with the {\em all} KK gauge bosons, including those
that belong to an unbroken symmetry, such as the KK gluons. This is due 
to the fact that the lightest KK gauge bosons are localized toward the IR 
brane, and are therefore strongly coupled to $t_L$, $b_L$ and $t_R$.  The 
FCNC interactions of KK gluons with $b_L$ also lead to contributions to 
hadronic $B$ decays, and they are potentially of the same order or even 
larger than the standard model gluonic penguins. These could result in 
sizeable deviations in CP asymmetries in $B$ decays such as $B \to \phi K_s$,
$B\to\eta' K_s$ and $B\to\pi^0 K_s$, among others~\cite{Burdman:2003nt}, even 
after the constrains from $Z \rightarrow b\bar{b}$ are taken into account. 

Finally, the large flavor violating coupling of the top quark, particularly 
$t_R$, may lead to a large contribution to $D^0$-$\bar{D}^0$ mixing.  This 
has the contributions both from KK gluon and $Z$ exchanges and has the form 
\begin{equation}
  \Delta m_D \simeq 4\pi\alpha_s\, \frac{\chi(c_R)}{2 m_1^2}\, 
    \frac{(U^{tu*}_R\,U^{tc}_R)^2}{2m_D}\, 
  \langle D^0|(\bar c_R\gamma_\mu u_R)(\bar c_R\gamma^\mu u_R)|\bar{D}^0\rangle,
\label{dmix}
\end{equation}
for the KK gluon exchange.  Here, $U_R$ is the rotation matrix for 
right-handed up quarks, and $\chi(c_R)$ is a function of $c_R$ which 
gives the enhancement due to the strong coupling of the KK gluons to 
$t_R$.  For instance, for $c_R\simeq 0$ and small brane couplings, 
$\chi\simeq 16$.  To estimate the contribution to $\Delta m_D$, we need 
the quark rotation matrix elements.  If we take $U^{tu*}_R U^{tc}_R 
\simeq \sin^5\!\theta_C$, with $\sin\theta_C \simeq 0.2$ the Cabibbo 
angle, then the current experimental limit~\cite{Burdman:2003rs} 
on $\Delta m_D$ translates into $m_1 \simgt 2~{\rm TeV}$.%
\footnote{Unlike for $U_L$ and $D_L$, there is in principle no 
reason why $U_R$ must have such scaling with the Cabibbo angle.}
In the strong bulk coupling case, $\chi(c_R)$ can be enhanced and 
somewhat larger $c_R$ or smaller mixing angles may be required.
The contribution from $Z$ is generically the same order but somewhat 
smaller.  We thus find that the effect can be consistent with but 
naturally close to the current experimental limit.  Similar contributions 
come from the interactions of $t_L$, but they are typically smaller 
than those from $t_R$ because of larger values of $c$.

\section{Conclusions and Discussion}
\label{sec:concl}

We have studied theories of electroweak symmetry breaking without 
a Higgs boson.  The electroweak symmetry is broken by non-trivial 
dynamics of a gauge interaction $G$, whose gauge coupling and the number 
of ``colors'' are denoted as $\tilde{g}$ and $N$, respectively.  In 
conventional 4D technicolor, the theory is assumed to be weakly coupled at 
the UV --- the loop expansion parameter $\L \equiv \tilde{g}^2 N/16\pi^2$ 
is smaller than unity --- and electroweak symmetry breaking is triggered 
at the IR where the perturbative expansion breaks down, $\L \simeq 1$. 
This makes the theory intractable because we must sum up all contributions 
of the form $\L^n$ ($n \in {\bf Z}$) arising at the $n$-th order in 
perturbation theory, to compute quantities such as the electroweak 
oblique parameters $S$ and $T$.  Moreover, such a theory generically 
has the problem of generating realistic fermion masses without 
conflicting with the experimental constraints on flavor violation.

In this paper we have studied an alternative possibility of ``dynamical'' 
electroweak symmetry breaking, in which the parameter $\L$ is larger than 
unity at the UV.  In particular, we have concentrated on the case where 
$\L$ stays almost constant over a wide energy range above the electroweak 
scale, as indicated by the curve (b) in Fig.~\ref{fig:holo}.  In such a 
theory quantities such as $S$ and $T$ can in principle be calculable
because they are given as expansions in powers of $1/\L$ and $1/N$.  
With sufficiently large $\L$ and $N$, therefore, we expect to have 
a calculable theory of electroweak symmetry breaking.   This is actually 
the case for a certain theory of this type, where there is a dual 
description in terms of a 5D theory compactified on the truncated AdS 
space.  In this dual description, the electroweak symmetry is simply 
broken by boundary conditions imposed at a spacetime boundary (the IR 
brane).   As long as the theory is weakly coupled and the AdS curvature 
scale $k$ is sufficiently small -- which correspond in the 4D picture 
to have sufficiently large $\L$ and $N$ -- we can reliably compute 
$S$ and $T$.  Quark and lepton masses are also obtained relatively 
easily by putting these fields in the 5D bulk and giving them masses 
at the IR brane.

Because the theory is calculable, it is possible to make a reliable 
comparison between its predictions and experimental data.  We have 
explicitly computed the gauge contribution to the $S$ parameter in 
the simplest potentially realistic theory of the kind discussed above 
--- a 5D warped space theory with the electroweak symmetry broken by 
the boundary conditions at the IR brane, and with the custodial $SU(2)$ 
symmetry imposed on the $G$ sector encoded in the physics of the 5D bulk 
and the IR boundary.  The result can be written in the form $S = c_S 
N/\pi$ (or $S = c_S N^2/\pi$), where $c_S$ is a positive constant 
of order one and $N$ (or $N^2$) is given by $16\pi^2/(g_L^2+g_R^2)k$ 
with $g_L$ and $g_R$ representing the 5D gauge couplings of the bulk 
gauge groups $SU(2)_L$ and $SU(2)_R$, respectively.  This result has 
a striking similarity to the estimate of $S$ in technicolor theories. 
This is because the size of the prediction of the theory for a physical 
quantity $P$ is in general rather insensitive to the value of $\L$ 
--- expanding $P$ in powers of $1/N$ as $P = \sum_n f_n(\L) N^{-n}$, 
the dependence of the functions $f_n(\L)$ on $\L$ is rather mild: 
$f_n(\L) = O(1)$ for the entire range of $\L$.  This in turn implies 
that if we want to have a large value of $N$, or equivalently a large 
energy interval where the theory has a weakly coupled 5D gravitational 
description, the gauge contribution to $S$ must be canceled by some 
other negative contribution, which does not arise from the gauge or 
matter sector of the model.  While such a contribution may arise, perhaps, 
from the intrinsic structure of the theory, for example from the sector 
needed to stabilize the radius of the extra dimension, it will most likely 
require an extension of the model.  The amount of cancellation required 
is typically of order $10\%$, but if it is attained, we can have a weakly 
coupled, calculable theory of ``dynamical'' electroweak symmetry breaking.

Another interesting possibility is to have a value of $N$ close to unity, 
which is attained by making one or both of the 5D gauge couplings $g_L$ 
and $g_R$ large. In this case we loose the calculability of the theory, 
but we expect that the corrections to the electroweak oblique parameters 
are small and of order $1/\pi$.  Realistic quark and lepton masses will 
also be obtained without contradicting with bounds from flavor changing 
processes, as this property is expected to persist even as we make the 5D 
theory strongly coupled.  We therefore arrive at a potentially consistent 
picture of a dynamical theory of electroweak symmetry breaking --- our 
theory is obtained by taking the limit of strong 5D gauge couplings 
in a warped 5D theory, in which the electroweak symmetry is broken 
by boundary conditions imposed at the IR brane. 

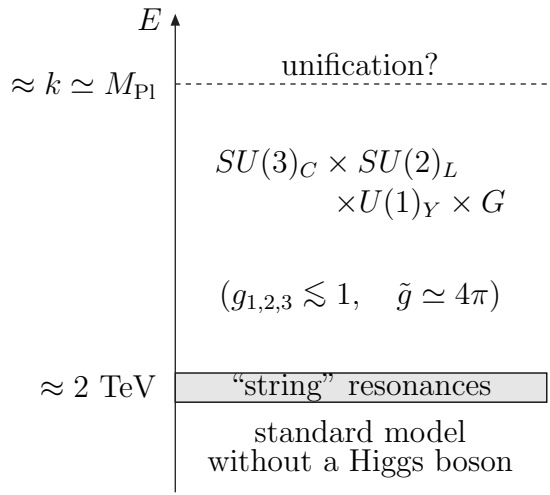
\begin{figure}[t]
\begin{center}
\begin{picture}(150,190)(-15,-26)
  \Text(70,4)[]{standard model} \Text(70,-8)[]{without a Higgs boson}
  \GBox(0,15)(140,26){0.9} \Text(70,21)[]{``string'' resonances}
  \DashLine(0,135)(140,135){2} \Text(70,139)[b]{unification?}
  \Text(15,105)[l]{$SU(3)_C \times SU(2)_L$}
  \Text(125,90)[r]{$\times U(1)_Y \times G$}
  \Text(70,55)[]{($g_{1,2,3} \simlt 1, \quad \tilde{g} \simeq 4\pi$)}
  \LongArrow(0,-19)(0,160) \Text(-5,160)[r]{$E$}
  \Text(-7,21)[r]{$\approx 2~{\rm TeV}$}
  \Text(-7,135)[r]{$\approx k \simeq M_{\rm Pl}$}
\end{picture}
\caption{The overall picture of the theory.  Here $g_{1,2,3}$ represents 
 the gauge couplings of $SU(3)_C \times SU(2)_L \times U(1)_Y$, while 
 $\tilde{g}$ represents that of $G$.} 
\label{fig:pic}
\end{center}
\end{figure}

A particularly interesting version of this theory is obtained by taking 
all the 5D gauge couplings to be strong:
\begin{equation}
  M_* \simeq \pi k \simeq \frac{16\pi^3}{g_{\rm 5D}^2},
\end{equation}
where $M_*$ is the cutoff scale of the theory and $g_{\rm 5D}$ represents 
all the 5D bulk gauge couplings $g_C$, $g_L$, $g_R$ and $g_X$ for 
$SU(3)_C$, $SU(2)_L$, $SU(2)_R$ and $U(1)_X$.  The observed gauge 
couplings then come almost entirely from the Planck-brane couplings.
The oblique parameters are expected to have a size 
\begin{equation}
  S,\, T \simlt \frac{1}{\pi},
\end{equation}
which is reasonably small, given the uncertainty of the estimate. 
Because $M_* \simeq \pi k$ implies that the IR cutoff scale $M'_*$ is 
close to the mass of the first KK resonance $m_1 \simeq \pi k'$, where 
$M'_* \equiv M_* e^{-\pi kR}$ and $k' \equiv k e^{-\pi kR}$, we do not 
have a weakly coupled KK picture around the TeV scale.  Rather, we have 
strongly coupled ``string states'' at the scale of
\begin{equation}
  M'_* \simeq 4\pi v \simeq 2~{\rm TeV},
\end{equation}
below which the theory is essentially the standard model without a Higgs 
boson (the radion field may also be lighter than $M'_*$).  A difference 
from the conventional 4D technicolor picture is that above this scale 
these states become associated with the 4D gauge interaction $G$ but 
whose gauge coupling $\tilde{g}$ stays almost constant and has a value 
of order $4\pi$ or slightly larger, i.e. $\kappa \simeq 1$ or slightly 
larger ($\L$ in 4D corresponds to $M_*/(\pi k)$ in 5D).  Just above 
$M'_*$, there are resonances, which then become increasingly broader 
and finally merge into a continuum consistent with conformal symmetry. 
This makes it possible to consider the theory as a strong coupling limit 
of a 5D warped theory.  We can then expect that realistic fermion masses 
are obtained without phenomenological disasters, although some flavor 
violating signals could be close to the experimental bounds.  This picture 
is also different from that of~\cite{Arkani-Hamed:1998rs}, because the 
theory as formulated in 4D is well defined up to the scale close to the 
4D Planck scale, $M_{\rm Pl}$.  The theory in the energy interval between 
$M'_*$ and $M_{\rm Pl}$ is simply $SU(3)_C \times SU(2)_L \times U(1)_Y 
\times G$ gauge theory with the coupling of $G$ nearly constant and of 
order $4\pi$.  For example, the running of the $SU(3)_C$, $SU(2)_L$ and 
$U(1)_Y$ gauge couplings is still logarithmic with the beta-function 
coefficients given by 
\begin{equation}
  b_i = b^{\rm SM}_i + \epsilon_i,
\end{equation}
where $i=3,2,1$ represents $SU(3)_C$, $SU(2)_L$ and $U(1)_Y$, and 
$b^{\rm SM}_i$ are the standard-model beta functions without a Higgs 
boson, $(b_1,b_2,b_3)=(4,-10/3,-7)$ [in the ``$SU(5)$ normalization'' 
for the $U(1)_Y$ gauge coupling]; $\epsilon_i$ are the corrections 
arising from the $G$ sector and $\epsilon_i \simeq 1$.  This may 
even suggest some sort of gauge unification at a high scale of order 
the Planck scale, since $\epsilon_i$ do not have an enhancement 
from the group theoretical factor so that the gross feature of the 
standard-model gauge coupling evolution (the three gauge couplings 
approach at high energies) is expected to persist.  The entire 
physical picture of our theory is depicted in Fig.~\ref{fig:pic}. 

The strong coupling feature of the theory raises the issue that the 5D 
gravitational description may not be entirely trustable.  For example, 
large quantum gravitational corrections may destabilize the structure of 
the background geometry.  However, we only need essentially the AdS-like 
structure to be preserved in the 5D bulk so that there remains a large 
energy interval above the electroweak scale where the theory is near 
conformal.  This may be the case, for example, when the 5D Planck 
scale $M_5$ is parametrically larger than the cutoff scale $M_*$. 
The full treatment of the theory will probably require string theoretic 
constructions, which may also give additional constraints: for example, 
the value of $N$ may be quantized through the Dirac quantization condition 
for higher-form gauge fields.  Nevertheless, given the presence of an 
effective field theoretic model with a certain strong coupling limit, 
it does not seem so implausible to expect that the theory as described 
here does in fact exist in some UV-completed schemes.

The experimental signatures of such a theory will be quite ``simple''. 
Below the scale of $\simeq 2~{\rm TeV}$, the theory is essentially the 
standard model without a Higgs boson.  New states appear at $M'_* \simeq 
2~{\rm TeV}$, which are composite states of the $G$ sector and will 
effectively be described as ``string'' states.  Since some of these 
states will be unstable, the tail of this physics may show up even at 
lower energies in collider experiments.  This situation is similar to 
the minimal technicolor theory, but here the observed fermion masses 
are correctly reproduced through physics at higher energies.  In this 
respect, it may not be easy to discriminate the present theory from 
certain technicolor models~\cite{Sundrum:1992xy} (they may even be 
related to each other in the space of $\L$).  An interesting state 
is the radion field which is expected to be lighter than $M'_*$, 
especially when $M_5 \simgt M_*$, and which arises in the 4D picture by 
spontaneous breaking of conformal invariance.  The properties of this 
field are similar to the standard model Higgs boson in some parameter 
region, but in general can be different~\cite{Giudice:2000av}.  Because 
the theory can tell quite little about physics at the $2~{\rm TeV}$ 
scale, it will be very important to explore this energy region 
experimentally.  Through such explorations, we will be able to learn 
about the physics of electroweak symmetry breaking caused by strong 
dynamics that does not contain any small parameter.

\vspace{0.2cm}
\begin{flushleft}
{\bf Note added:} \\
After the completion of this work, Ref.~\cite{Davoudiasl:2003me} 
appeared which addresses related issues.
\end{flushleft}

\section*{Acknowledgments}

This work was supported in part by the Director, Office of Science, 
Office of High Energy and Nuclear Physics of the U.S. Department of 
Energy under Contract DE-AC0376SF00098.

\newpage

\section*{Appendix~A}

In this appendix we calculate the oblique parameters $S$, $T$ and $U$ at 
the leading order for the theory described in section~\ref{sec:model}. 
These parameters can be calculated by integrating out the new physics, 
i.e. the $G$ sector, and deriving the low-energy effective theory for 
the electroweak gauge bosons.  The effects of the $G$ sector then appear 
in the vacuum polarizations for these gauge fields, which we parameterize 
by $\Pi_{XY}$ and $\Pi'_{XY}$ ($X,Y=1,3,Q$) as
\begin{eqnarray}
  {\cal L}_{\rm eff} &=& -\frac{1}{2} \sum_{a=1}^{2} 
    W^a_\mu \Biggl[ \Bigl( \frac{v^2}{2}+\Pi_{11} \Bigr)
      + p^2 \Bigl( \frac{1}{g^2}-\Pi'_{11} \Bigr) \Biggr] W^a_\mu
\nonumber\\
  && -\frac{1}{2} W^3_\mu \Biggl[ \Bigl( \frac{v^2}{2}+\Pi_{33} \Bigr)
      + p^2 \Bigl( \frac{1}{g^2}-\Pi'_{33} \Bigr) \Biggr] W^3_\mu
\nonumber\\
  && -\frac{1}{2} B_\mu \Biggl[ \Bigl( \frac{v^2}{2}+\Pi_{33} \Bigr)
      + p^2 \Bigl( \frac{1}{g'^2}-\Pi'_{33}+2\Pi'_{3Q}-\Pi'_{QQ} 
      \Bigr) \Biggr] B_\mu
\nonumber\\
  && + W^3_\mu \Biggl[ \Bigl( \frac{v^2}{2}+\Pi_{33} \Bigr)
      - p^2 \Bigl( \Pi'_{33}-\Pi'_{3Q} \Bigr) \Biggr] B_\mu,
\label{eq:def-Pi}
\end{eqnarray}
where ${\cal L}_{\rm eff}$ is the low-energy effective Lagrangian 
in the 4D momentum space.  The normalizations for the gauge fields 
are taken such that they couple to the matter fields through the 
covariant derivatives
\begin{eqnarray}
  {\cal D}_\mu \psi_L &=& \partial_\mu \psi_L
    + \frac{i}{2} \pmatrix{W^3_\mu+2YB_\mu & W^1_\mu-iW^2_\mu 
      \cr W^1_\mu+iW^2_\mu & -W^3_\mu+2YB_\mu} \psi_L,
\\
  {\cal D}_\mu \psi_R &=& \partial_\mu \psi_R
    + iY B_\mu \psi_R,
\end{eqnarray}
where $\psi_L$ and $\psi_R$ represent $\{ q,l \}$ and $\{ u,d,e \}$, 
respectively.  The parameters $S$, $T$ and $U$ are defined 
by~\cite{Peskin:1990zt}
\begin{eqnarray}
  S &\equiv& 16\pi ( \Pi'_{33} - \Pi'_{3Q} ),
\\
  T &\equiv& \frac{8\pi(g^2+g'^2)}{g^2g'^2v^2} 
    ( \Pi_{11} - \Pi_{33} ),
\\
  U &\equiv& 16\pi ( \Pi'_{11} - \Pi'_{33} ).
\end{eqnarray}
These parameters do not depend on the values of $g$ and $g'$ we choose 
to extract $\Pi$s and $\Pi'$s from Eq.~(\ref{eq:def-Pi}), as long as 
$\Pi_{11,33} \ll v^2$ and $\Pi'_{11,33,3Q,QQ} \ll 1/g^2,1/g'^2$.

The effective Lagrangian, ${\cal L}_{\rm eff}$, in our theory is obtained 
in 5D by integrating out the physics of $y > 0$ keeping the values of 
the 5D gauge fields at $y=0$ fixed, which we identify as the low-energy 
4D fields~\cite{Barbieri:2003pr}.  Since we are here interested in the 
leading-order contributions to $S$, $T$ and $U$, which are represented 
in the 4D picture by the diagram shown in Fig.~\ref{fig:diag-1} in 
section~\ref{sec:structure}, it is sufficient to solve the equations 
of motion for the bulk gauge fields at the classical level under 
appropriate boundary conditions. 

The action for the gauge fields are given by Eqs.~(\ref{eq:gauge-kinetic},%
~\ref{eq:TeV-gauge-kinetic}), i.e. the sum of the bulk, Planck-brane, 
and TeV-brane gauge kinetic terms.  In terms of the conformal coordinate 
$z \equiv e^{ky}/k$, the bulk equations of motion for the gauge fields 
are written as
\begin{equation}
  z\, \partial_z \Bigl( \frac{1}{z} \partial_z A^G_\mu \Bigr) 
    - p^2 A^G_\mu = 0,
\end{equation}
where we have kept only the transverse modes, and $G$ runs for $L1, R1, 
L3, R3$, and $X$, which represent the first two components of $SU(2)_L$, 
those of $SU(2)_R$, the third component of $SU(2)_L$, that of $SU(2)_R$, 
and $U(1)_X$, respectively.  These equations have solutions of the form
\begin{equation}
  A^G_\mu(p,z) = z \Bigl\{ a^G_\mu(p) I_1(pz) + b^G_\mu(p) K_1(pz) \Bigr\},
\label{eq:ap-sol}
\end{equation}
where $I_1(x)$ and $K_1(x)$ are the modified Bessel functions, and 
$a^G_\mu(p)$ and $b^G_\mu(p)$ are functions of the 4D momentum $p$. 
The boundary conditions at the Planck brane ($y=0$) are given by
\begin{eqnarray}
  \Bigl[ A^{L1}_\mu(p,z) \Bigr]_{z=\frac{1}{k}} &=& W^1_\mu(p),
\label{eq:ap-bc-1} \\
  \Bigl[ A^{R1}_\mu(p,z) \Bigr]_{z=\frac{1}{k}} &=& 0,
\label{eq:ap-bc-2} \\
  \Bigl[ A^{L3}_\mu(p,z) \Bigr]_{z=\frac{1}{k}} &=& W^3_\mu(p),
\label{eq:ap-bc-3} \\
  \Bigl[ A^{R3}_\mu(p,z) \Bigr]_{z=\frac{1}{k}} &=& B_\mu(p),
\label{eq:ap-bc-4} \\
  \Bigl[ A^{X}_\mu(p,z) \Bigr]_{z=\frac{1}{k}}  &=& B_\mu(p),
\label{eq:ap-bc-5}
\end{eqnarray}
which identify the 5D fields at the Planck brane to the low-energy 
4D degrees of freedom.  The boundary conditions at the TeV brane 
($y=\pi R$) are given by 
\begin{eqnarray}
  && \Bigl[ A^{L1}_\mu(p,z) - A^{R1}_\mu(p,z) \Bigr]_{z=\frac{1}{k'}} = 0,
\label{eq:ap-bc-6} \\
  && \Bigl[ \frac{1}{g_L^2} \partial_z A^{L1}_\mu(p,z) 
    + \frac{1}{g_R^2} \partial_z A^{R1}_\mu(p,z) 
    + p^2 \frac{k}{k'}\Bigl( \tilde{Z}_L A^{L1}_\mu(p,z) 
      + \tilde{Z}_R A^{R1}_\mu(p,z) \Bigr) \Bigr]_{z=\frac{1}{k'}} = 0,
\label{eq:ap-bc-7} \\
  && \Bigl[ A^{L3}_\mu(p,z) - A^{R3}_\mu(p,z) \Bigr]_{z=\frac{1}{k'}} = 0,
\label{eq:ap-bc-8} \\
  && \Bigl[ \frac{1}{g_L^2} \partial_z A^{L3}_\mu(p,z) 
    + \frac{1}{g_R^2} \partial_z A^{R3}_\mu(p,z) 
    + p^2 \frac{k}{k'}\Bigl( \tilde{Z}_L A^{L3}_\mu(p,z) 
      + \tilde{Z}_R A^{R3}_\mu(p,z) \Bigr) \Bigr]_{z=\frac{1}{k'}} = 0,
\label{eq:ap-bc-9} \\
  && \Bigl[ \frac{1}{g_X^2} \partial_z A^X_\mu(p,z) 
    + p^2 \frac{k}{k'} Z_X A^X_\mu(p,z) \Bigr]_{z=\frac{1}{k'}} = 0,
\label{eq:ap-bc-10}
\end{eqnarray}
which are essentially those of Eq.~(\ref{eq:bc-pi}) but appropriately 
modified by the presence of the TeV-brane gauge kinetic terms.  Here, 
$\tilde{Z}_L \equiv Z_L+Z_M$ and $\tilde{Z}_R \equiv Z_R+Z_M$.

The boundary conditions Eqs.~(\ref{eq:ap-bc-1}~--~\ref{eq:ap-bc-10}) 
determine the coefficients $a^G_\mu(p)$ and $b^G_\mu(p)$ in 
Eq.~(\ref{eq:ap-sol}).  Plugging these solutions into the original action,
 Eq.~(\ref{eq:gauge-kinetic}) and Eq.~(\ref{eq:TeV-gauge-kinetic}), and 
integrating over $z$, we obtain the low-energy effective Lagrangian.
Expanding this Lagrangian in powers of $p$ up to the quadratic order, 
we find 
\begin{eqnarray}
  {\cal L}_{\rm eff} &=& 
    -\frac{k'^2}{(g_L^2+g_R^2)k} \Bigl( \sum_{a=1}^{2} W^a_\mu W^a_\mu 
    + W^3_\mu W^3_\mu - 2 W^3_\mu B_\mu + B_\mu B_\mu \Bigr)
\nonumber\\
  && -\frac{p^2}{2} \Biggl\{ 
    \Bigl( \frac{\pi R}{g_L^2}+\frac{1}{\tilde{g}_L^2} \Bigr)
    \Bigl( \sum_{a=1}^{2} W^a_\mu W^a_\mu + W^3_\mu W^3_\mu \Bigr)
    +\Bigl( \frac{\pi R}{g_R^2}+\frac{\pi R}{g_X^2}
    +\frac{1}{\tilde{g}_Y^2} \Bigr) B_\mu B_\mu \Biggr\}
\nonumber\\
  && -\frac{p^2}{8(g_L^2+g_R^2)k} \Biggl\{ 
    \Bigl( -3 + \frac{4g_R^4k}{g_L^2+g_R^2}(\tilde{Z}_L+\tilde{Z}_R) 
    \Bigr) \Bigl( \sum_{a=1}^{2} W^a_\mu W^a_\mu + W^3_\mu W^3_\mu \Bigr)
\nonumber\\
  && \qquad\qquad\qquad
    + \Bigl( -3 + \frac{4g_L^4k}{g_L^2+g_R^2}(\tilde{Z}_L+\tilde{Z}_R) 
    + 4(g_L^2+g_R^2)k Z_X \Bigr) B_\mu B_\mu
\nonumber\\
  && \qquad\qquad\qquad
    + \Bigl( 6 + \frac{8g_L^2g_R^2k}{g_L^2+g_R^2}(\tilde{Z}_L+\tilde{Z}_R) 
    \Bigr) W^3_\mu B_\mu \Biggr\}.
\end{eqnarray}
This reproduces the matching relations of Eqs.~(\ref{eq:corresp}) at the 
leading order in $1/\pi kR$ (assuming $Z_{L,R,M,X} = O(1/16\pi^2)$ as 
suggested by NDA).  We then obtain the vacuum polarization parameters
\begin{eqnarray}
  && \Pi_{11} = \Pi_{33} = 0,
\label{eq:vac-pol-1} \\
  && \Pi'_{11} = \Pi'_{33} = \frac{3}{4(g_L^2+g_R^2)k}
     - \frac{g_R^4}{(g_L^2+g_R^2)^2}(\tilde{Z}_L+\tilde{Z}_R),
\label{eq:vac-pol-2} \\
  && \Pi'_{3Q} = - \frac{g_R^2}{g_L^2+g_R^2}(\tilde{Z}_L+\tilde{Z}_R),
\label{eq:vac-pol-3} \\
  && \Pi'_{QQ} = -(\tilde{Z}_L+\tilde{Z}_R+Z_X).
\label{eq:vac-pol-4}
\end{eqnarray}
Several features of this result can be understood from the symmetry 
reason.  For example, the reason for why $\Pi_{11} = \Pi_{33}$ and 
$\Pi'_{11} = \Pi'_{33}$ comes from the fact that the dynamics of the 
$G$ sector, encoded in the bulk and TeV-brane physics, respects the 
custodial $SU(2)$ symmetry.  The coefficients for TeV-brane operators 
$Z_L$, $Z_R$ and $Z_M$ always appear in the combination of $\tilde{Z}_L 
+ \tilde{Z}_R = Z_L + Z_R + Z_M$ because the TeV brane respects the 
diagonal subgroup of $SU(2)_L$ and $SU(2)_R$.  Finally, in the case 
of vanishing TeV-brane operators, $Z_L = Z_R = Z_M = Z_X =0$, we get 
$\Pi'_{3Q} =  \Pi'_{QQ} = 0$.  This is because the full $SU(2)_L 
\times SU(2)_R$ symmetry is respected by the operators in the 5D bulk. 

Eqs.~(\ref{eq:vac-pol-1}~--~\ref{eq:vac-pol-4}) give the oblique 
parameters
\begin{eqnarray}
  && S = \frac{16\pi}{(g_L^2+g_R^2)k}
    \Biggl\{ \frac{3}{4} + \frac{g_L^2k \cdot g_R^2k}
    {(g_L^2+g_R^2)k}(\tilde{Z}_L+\tilde{Z}_R) \Biggr\},
\\
  && T = U = 0.
\end{eqnarray}
As expected, the $T$ and $U$ parameters are zero at this order 
because of the custodial $SU(2)$ symmetry imposed on the $G$ sector. 
The value of $S$ has a size of order $N/\pi$, where $N$ is given by 
Eq.~(\ref{eq:def-N}), as discussed in section~\ref{subsec:structure-1}.

\section*{Appendix~B}

In this appendix we derive the formula determining the mass eigenvalues 
for the quark and lepton KK towers.  We here use the notation of 
Eq.~(\ref{eq:quark-rep}) for the up-type quark sector.  However, the 
computation is completely identical for the down-type quark and lepton 
sectors, so that the results are also applicable to these cases. 

We define the rescaled fields $\hat{u} = e^{-2ky} u$, $\hat{u}^c = 
e^{-2ky} u^c$, $\hat{\bar{u}} = e^{-2ky} \bar{u}$, and $\hat{\bar{u}}^c 
= e^{-2ky} \bar{u}^c$.  In terms of these fields, the action is 
written as 
\begin{eqnarray}
  {\cal S} &=& \int\!d^4x \int\!dy\, \Biggl[ e^{ky}
    \Bigl( \hat{u}^\dagger i \bar{\sigma}^\mu \partial_\mu \hat{u}
      + \hat{u}^c i \sigma^\mu \partial_\mu \hat{u}^{c\dagger}
      + \hat{\bar{u}} i \sigma^\mu \partial_\mu \hat{\bar{u}}^\dagger
      + \hat{\bar{u}}^{c\dagger} i \bar{\sigma}^\mu \partial_\mu 
        \hat{\bar{u}}^c \Bigr)
\nonumber\\
  && + \hat{u}^c (\partial_y + c_L k) \hat{u}
     + \hat{u}^\dagger (-\partial_y + c_L k) \hat{u}^{c\dagger}
     + \hat{\bar{u}}^c (\partial_y + c_R k) \hat{\bar{u}} 
     + \hat{\bar{u}}^\dagger (-\partial_y + c_R k) \hat{\bar{u}}^{c\dagger} 
\nonumber\\
  && - \delta(y-\pi R) 
     \Bigl( \lambda \hat{u} \hat{\bar{u}}
     + \lambda^* \hat{u}^\dagger \hat{\bar{u}}^\dagger \Bigr) \Biggr],
\label{eq:top-kinetic}
\end{eqnarray}
where the bulk terms come from the 5D kinetic terms and the TeV-brane 
terms from the operator in Eq.~(\ref{eq:yukawa}); $c_L$ and $c_R$ 
are the bulk mass parameters for the $q$ and $\psi_{\bar{u}}$ fields, 
i.e. $c_L = c_q$ and $c_R = c_{\psi_{\bar{u}}}$, and $\lambda$ 
is given by $\lambda = y_u v_H$.

The above action, Eq.~(\ref{eq:top-kinetic}), provides both bulk 
equations of motion and boundary conditions.  Expanding the 5D 
fields as
\begin{eqnarray}
  \hat{u}(x,y)         &=& u(x)       \hat{f}_u(y),          \\
  \hat{u}^c(x,y)       &=& \bar{u}(x) \hat{f}_u^c(y),        \\
  \hat{\bar{u}}(x,y)   &=& \bar{u}(x) \hat{f}_{\bar{u}}(y),  \\
  \hat{\bar{u}}^c(x,y) &=& u(x)       \hat{f}_{\bar{u}}^c(y),
\end{eqnarray}
we obtain the bulk equations of motion
\begin{eqnarray}
  && \Bigl( -\partial_z + \frac{c_L}{z} \Bigr) f_u^c 
    + m f_u = 0, 
\label{eq:bulk-eom-1} \\
  && \Bigl( \partial_z + \frac{c_L}{z} \Bigr) f_u 
    + m f_u^c = 0, 
\label{eq:bulk-eom-2} \\
  && \Bigl( -\partial_z + \frac{c_R}{z} \Bigr) f_{\bar{u}}^c 
    + m f_{\bar{u}} = 0, 
\label{eq:bulk-eom-3} \\
  && \Bigl( \partial_z + \frac{c_R}{z} \Bigr) f_{\bar{u}} 
    + m f_{\bar{u}}^c = 0, 
\label{eq:bulk-eom-4} 
\end{eqnarray}
where $m$ represents the 4D mass eigenvalues, $i\bar{\sigma}^\mu 
\partial_\mu q = m \bar{q}^\dagger$.  Here, we have used 
the conformal coordinate $z \equiv e^{ky}/k$, and $f_u(z) 
= \hat{f}_u(\ln(kz)/k)$, $f_u^c = \hat{f}_u^c(\ln(kz)/k)$, 
$f_{\bar{u}} = \hat{f}_{\bar{u}}(\ln(kz)/k)$ and $f_{\bar{u}}^c 
= \hat{f}_{\bar{u}}^c(\ln(kz)/k)$.  The boundary conditions 
are given by
\begin{equation}
  \left\{ \begin{array}{l}
    f_u^c\Bigr|_{z = \frac{1}{k}+\epsilon} = 0, \\
    (\partial_z + k c_L) f_u\Bigr|_{z = \frac{1}{k}+\epsilon} = 0, \\
    f_{\bar{u}}^c\Bigr|_{z = \frac{1}{k}+\epsilon} = 0, \\
    (\partial_z + k c_R) f_{\bar{u}}\Bigr|_{z = \frac{1}{k}+\epsilon} = 0,
  \end{array} \right. 
\qquad
  \left\{ \begin{array}{l}
    f_u^c\Bigr|_{z = \frac{1}{k'}-\epsilon} 
      - \lambda f_{\bar{u}}|_{z = \frac{1}{k'}-\epsilon} = 0, \\
    (\partial_z + k' c_L) f_u\Bigr|_{z = \frac{1}{k'}-\epsilon} 
      + m f_u^c\Bigr|_{z = \frac{1}{k'}-\epsilon} = 0, \\
    f_{\bar{u}}^c\Bigr|_{z = \frac{1}{k'}-\epsilon} 
      - \lambda f_u|_{z = \frac{1}{k'}-\epsilon} = 0, \\
     (\partial_z + k' c_R) f_{\bar{u}}\Bigr|_{z = \frac{1}{k'}-\epsilon} 
      + m f_{\bar{u}}^c\Bigr|_{z = \frac{1}{k'}-\epsilon} = 0,
  \end{array} \right.
\label{eq:top-bc}
\end{equation}
where $\epsilon \rightarrow 0$.  In Eqs.~(\ref{eq:bulk-eom-1}~--~%
\ref{eq:top-bc}), we have rotated the phases of the fields such that 
the coupling $\lambda$, and thus the 4D masses $m$, becomes real.

Eqs.~(\ref{eq:bulk-eom-1}~--~\ref{eq:bulk-eom-4}) have the solutions 
of the form
\begin{eqnarray}
  f_u &=&
    \sqrt{z}\, \Bigl\{ a_u J_{c_L+\frac{1}{2}}(mz) 
    + b_u Y_{c_L+\frac{1}{2}}(mz) \Bigr\},
\\
  f_u^c &=&
    \sqrt{z}\, \Bigl\{ a_u^c J_{c_L-\frac{1}{2}}(mz) 
    + b_u^c Y_{c_L-\frac{1}{2}}(mz) \Bigr\},
\\
  f_{\bar{u}} &=&
    \sqrt{z}\, \Bigl\{ a_{\bar{u}} J_{c_R+\frac{1}{2}}(mz) 
    + b_{\bar{u}} Y_{c_R+\frac{1}{2}}(mz) \Bigr\},
\\
  f_{\bar{u}}^c &=&
    \sqrt{z}\, \Bigl\{ a_{\bar{u}}^c J_{c_R-\frac{1}{2}}(mz) 
    + b_{\bar{u}}^c Y_{c_R-\frac{1}{2}}(mz) \Bigr\},
\end{eqnarray}
where $a_u$, $b_u$, $a_u^c$, $b_u^c$, $a_{\bar{u}}$, $b_{\bar{u}}$, 
$a_{\bar{u}}^c$ and $b_{\bar{u}}^c$ are constants.  These constants 
are determined by the boundary conditions, Eq.~(\ref{eq:top-bc}). 
Non-trivial solutions are then obtained only when the following 
relation is satisfied:
\begin{eqnarray}
  && \Biggl( J_{c_L-\frac{1}{2}}\Bigl(\frac{m}{k'}\Bigl) 
    - \frac{J_{c_L-\frac{1}{2}}\bigl(\frac{m}{k}\bigl)}
      {Y_{c_L-\frac{1}{2}}\bigl(\frac{m}{k}\bigl)}
      Y_{c_L-\frac{1}{2}}\Bigl(\frac{m}{k'}\Bigl) \Biggr)
  \Biggl( J_{c_R-\frac{1}{2}}\Bigl(\frac{m}{k'}\Bigl) 
    - \frac{J_{c_R-\frac{1}{2}}\bigl(\frac{m}{k}\bigl)}
      {Y_{c_R-\frac{1}{2}}\bigl(\frac{m}{k}\bigl)}
      Y_{c_R-\frac{1}{2}}\Bigl(\frac{m}{k'}\Bigl) \Biggr)
\nonumber\\
  && - \lambda^2 
  \Biggl( J_{c_L+\frac{1}{2}}\Bigl(\frac{m}{k'}\Bigl) 
    - \frac{J_{c_L-\frac{1}{2}}\bigl(\frac{m}{k}\bigl)}
      {Y_{c_L-\frac{1}{2}}\bigl(\frac{m}{k}\bigl)}
      Y_{c_L+\frac{1}{2}}\Bigl(\frac{m}{k'}\Bigl) \Biggr)
  \Biggl( J_{c_R+\frac{1}{2}}\Bigl(\frac{m}{k'}\Bigl) 
    - \frac{J_{c_R-\frac{1}{2}}\bigl(\frac{m}{k}\bigl)}
      {Y_{c_R-\frac{1}{2}}\bigl(\frac{m}{k}\bigl)}
      Y_{c_R+\frac{1}{2}}\Bigl(\frac{m}{k'}\Bigl) \Biggr) = 0.
\end{eqnarray}
This is the equation cited in the text as Eq.~(\ref{eq:det-mass}). 
The mass eigenvalues, $m$, are determined as solutions of this 
equation. 

For the down-type quarks, we have to use $c_L = c_q$, 
$c_R = c_{\psi_{\bar{d}}}$ and $\lambda = y_d v_H$, instead of 
$c_L = c_q$, $c_R = c_{\psi_{\bar{u}}}$ and $\lambda = y_u v_H$.
For the leptons, $c_L = c_l$, $c_R = c_{\psi_{\bar{e}}}$ 
and $\lambda = y_e v_H$.


\newpage

\begin{thebibliography}{99}

\bibitem{Weinberg:gm}
S.~Weinberg,
Phys.\ Rev.\ D {\bf 13}, 974 (1976);
Phys.\ Rev.\ D {\bf 19}, 1277 (1979);
L.~Susskind,
Phys.\ Rev.\ D {\bf 20}, 2619 (1979).

\bibitem{Kaplan:1983fs}
D.~B.~Kaplan and H.~Georgi,
Phys.\ Lett.\ B {\bf 136}, 183 (1984);
D.~B.~Kaplan, H.~Georgi and S.~Dimopoulos,
Phys.\ Lett.\ B {\bf 136}, 187 (1984).

\bibitem{Maldacena:1997re}
J.~M.~Maldacena,
Adv.\ Theor.\ Math.\ Phys.\  {\bf 2}, 231 (1998)
[Int.\ J.\ Theor.\ Phys.\  {\bf 38}, 1113 (1999)]
[arXiv:hep-th/9711200];
S.~S.~Gubser, I.~R.~Klebanov and A.~M.~Polyakov,
Phys.\ Lett.\ B {\bf 428}, 105 (1998)
[arXiv:hep-th/9802109];
E.~Witten,
Adv.\ Theor.\ Math.\ Phys.\  {\bf 2}, 253 (1998)
[arXiv:hep-th/9802150].

\bibitem{Contino:2003ve}
R.~Contino, Y.~Nomura and A.~Pomarol,
Nucl.\ Phys.\ B {\bf 671}, 148 (2003)
[arXiv:hep-ph/0306259];
T.~Gherghetta and A.~Pomarol,
Phys.\ Rev.\ D {\bf 67}, 085018 (2003)
[arXiv:hep-ph/0302001].

\bibitem{Csaki:2003zu}
C.~Csaki, C.~Grojean, L.~Pilo and J.~Terning,
arXiv:hep-ph/0308038.

\bibitem{Nomura:2003du}
Y.~Nomura,
JHEP {\bf 0311}, 050 (2003)
[arXiv:hep-ph/0309189].

\bibitem{Csaki:2003dt}
C.~Csaki, C.~Grojean, H.~Murayama, L.~Pilo and J.~Terning,
arXiv:hep-ph/0305237.

\bibitem{Barbieri:2003pr}
R.~Barbieri, A.~Pomarol and R.~Rattazzi,
arXiv:hep-ph/0310285.

\bibitem{Peskin:1990zt}
M.~E.~Peskin and T.~Takeuchi,
Phys.\ Rev.\ Lett.\  {\bf 65}, 964 (1990);
Phys.\ Rev.\ D {\bf 46}, 381 (1992).

\bibitem{Randall:1999ee}
L.~Randall and R.~Sundrum,
Phys.\ Rev.\ Lett.\ {\bf 83}, 3370 (1999)
[arXiv:hep-ph/9905221].

\bibitem{Dimopoulos:1979es}
S.~Dimopoulos and L.~Susskind,
Nucl.\ Phys.\ B {\bf 155}, 237 (1979);
E.~Eichten and K.~D.~Lane,
Phys.\ Lett.\ B {\bf 90}, 125 (1980).

\bibitem{Holdom:1981rm}
B.~Holdom,
Phys.\ Rev.\ D {\bf 24}, 1441 (1981);
K.~Yamawaki, M.~Bando and K.~i.~Matumoto,
Phys.\ Rev.\ Lett.\  {\bf 56}, 1335 (1986);
T.~W.~Appelquist, D.~Karabali and L.~C.~R.~Wijewardhana,
Phys.\ Rev.\ Lett.\  {\bf 57}, 957 (1986).

\bibitem{'tHooft:1973jz}
G.~'t Hooft,
Nucl.\ Phys.\ B {\bf 72}, 461 (1974).

\bibitem{Nomura:2001mf}
Y.~Nomura, D.~R.~Smith and N.~Weiner,
Nucl.\ Phys.\ B {\bf 613}, 147 (2001)
[arXiv:hep-ph/0104041].

\bibitem{Goldberger:2002pc}
W.~D.~Goldberger, Y.~Nomura and D.~R.~Smith,
Phys.\ Rev.\ D {\bf 67}, 075021 (2003)
[arXiv:hep-ph/0209158];
S.~J.~Huber and Q.~Shafi,
arXiv:hep-ph/0309252.

\bibitem{Arkani-Hamed:2001is}
N.~Arkani-Hamed, A.~G.~Cohen and H.~Georgi,
Phys.\ Lett.\ B {\bf 516}, 395 (2001)
[arXiv:hep-th/0103135];
T.~Hirayama and K.~Yoshioka,
arXiv:hep-th/0311233.

\bibitem{Gherghetta:2000qt}
T.~Gherghetta and A.~Pomarol,
Nucl.\ Phys.\ B {\bf 586}, 141 (2000)
[arXiv:hep-ph/0003129];
S.~J.~Huber and Q.~Shafi,
Phys.\ Lett.\ B {\bf 498}, 256 (2001)
[arXiv:hep-ph/0010195].

\bibitem{Goldberger:2002cz}
W.~D.~Goldberger and I.~Z.~Rothstein,
Phys.\ Rev.\ Lett.\ {\bf 89}, 131601 (2002)
[arXiv:hep-th/0204160];
arXiv:hep-th/0208060;
Y.~Nomura and D.~R.~Smith,
Phys.\ Rev.\ D {\bf 68}, 075003 (2003)
[arXiv:hep-ph/0305214].

\bibitem{Manohar:1983md}
A.~Manohar and H.~Georgi,
Nucl.\ Phys.\ B {\bf 234}, 189 (1984);
H.~Georgi and L.~Randall,
Nucl.\ Phys.\ B {\bf 276}, 241 (1986);
Z.~Chacko, M.~A.~Luty and E.~Ponton,
JHEP {\bf 0007}, 036 (2000)
[arXiv:hep-ph/9909248];
Y.~Nomura,
Phys.\ Rev.\ D {\bf 65}, 085036 (2002)
[arXiv:hep-ph/0108170].

\bibitem{Arkani-Hamed:2000ds}
N.~Arkani-Hamed, M.~Porrati and L.~Randall,
JHEP {\bf 0108}, 017 (2001)
[arXiv:hep-th/0012148];
R.~Rattazzi and A.~Zaffaroni,
JHEP {\bf 0104}, 021 (2001)
[arXiv:hep-th/0012248];
M.~Perez-Victoria,
JHEP {\bf 0105}, 064 (2001)
[arXiv:hep-th/0105048].

\bibitem{Witten:1979kh}
See, for example, 
E.~Witten,
Nucl.\ Phys.\ B {\bf 160}, 57 (1979).

\bibitem{Agashe:2003zs}
K.~Agashe, A.~Delgado, M.~J.~May and R.~Sundrum,
JHEP {\bf 0308}, 050 (2003)
[arXiv:hep-ph/0308036].

\bibitem{Georgi:1991ci}
H.~Georgi,
Nucl.\ Phys.\ B {\bf 363}, 301 (1991);
M.~J.~Dugan and L.~Randall,
Phys.\ Lett.\ B {\bf 264}, 154 (1991);
E.~Gates and J.~Terning,
Phys.\ Rev.\ Lett.\  {\bf 67}, 1840 (1991);
M.~A.~Luty and R.~Sundrum,
Phys.\ Rev.\ Lett.\  {\bf 70} (1993) 529
[arXiv:hep-ph/9209255].

\bibitem{Holdom:1990xp}
B.~Holdom,
Phys.\ Lett.\ B {\bf 259}, 329 (1991);
P.~Langacker and M.~x.~Luo,
Phys.\ Rev.\ D {\bf 45}, 278 (1992).

\bibitem{Csaki:2000zn}
C.~Csaki, M.~L.~Graesser and G.~D.~Kribs,
Phys.\ Rev.\ D {\bf 63}, 065002 (2001)
[arXiv:hep-th/0008151];
J.~F.~Gunion, M.~Toharia and J.~D.~Wells,
arXiv:hep-ph/0311219.

\bibitem{Hall:2001tn}
L.~J.~Hall, H.~Murayama and Y.~Nomura,
Nucl.\ Phys.\ B {\bf 645}, 85 (2002)
[arXiv:hep-th/0107245];
R.~Sekhar Chivukula, D.~A.~Dicus and H.~J.~He,
Phys.\ Lett.\ B {\bf 525}, 175 (2002)
[arXiv:hep-ph/0111016];
Y.~Abe, N.~Haba, Y.~Higashide, K.~Kobayashi and M.~Matsunaga,
Prog.\ Theor.\ Phys.\  {\bf 109}, 831 (2003)
[arXiv:hep-th/0302115].

\bibitem{Csaki:2003sh}
C.~Csaki, C.~Grojean, J.~Hubisz, Y.~Shirman and J.~Terning,
arXiv:hep-ph/0310355.

\bibitem{Henningson:1998cd}
M.~Henningson and K.~Sfetsos,
Phys.\ Lett.\ B {\bf 431}, 63 (1998)
[arXiv:hep-th/9803251];
W.~Muck and K.~S.~Viswanathan,
Phys.\ Rev.\ D {\bf 58}, 106006 (1998)
[arXiv:hep-th/9805145];
J.~Garriga and A.~Pomarol,
Phys.\ Lett.\ B {\bf 560}, 91 (2003)
[arXiv:hep-th/0212227].

\bibitem{Abe:2001nn}
See, for example,
T.~Abe {\it et al.}  [American Linear Collider Working Group Collaboration],
in {\it Proc. of the APS/DPF/DPB Summer Study on the Future of Particle Physics (Snowmass 2001) } ed. N.~Graf,
arXiv:hep-ex/0106055;
arXiv:hep-ex/0106057.

\bibitem{Atwood:2003tg}
D.~Atwood and G.~Hiller,
arXiv:hep-ph/0307251.

\bibitem{Agashe}
K.~Agashe,
Talk given at the Second Workshop on the Discovery Potential 
of an Asymmetric B Factory at $10^{36}$ Luminosity,
SLAC, Oct 2003.

\bibitem{Burdman:2003nt}
G.~Burdman,
arXiv:hep-ph/0310144.

\bibitem{Burdman:2003rs}
G.~Burdman and I.~Shipsey,
arXiv:hep-ph/0310076.

\bibitem{Arkani-Hamed:1998rs}
N.~Arkani-Hamed, S.~Dimopoulos and G.~R.~Dvali,
Phys.\ Lett.\ B {\bf 429}, 263 (1998)
[arXiv:hep-ph/9803315];
I.~Antoniadis, N.~Arkani-Hamed, S.~Dimopoulos and G.~R.~Dvali,
Phys.\ Lett.\ B {\bf 436}, 257 (1998)
[arXiv:hep-ph/9804398].

\bibitem{Sundrum:1992xy}
For example, 
R.~Sundrum,
Nucl.\ Phys.\ B {\bf 395}, 60 (1993)
[arXiv:hep-ph/9205203].

\bibitem{Giudice:2000av}
G.~F.~Giudice, R.~Rattazzi and J.~D.~Wells,
Nucl.\ Phys.\ B {\bf 595}, 250 (2001)
[arXiv:hep-ph/0002178];
C.~Csaki, M.~L.~Graesser and G.~D.~Kribs,
in~\cite{Csaki:2000zn}.

\bibitem{Davoudiasl:2003me}
H.~Davoudiasl, J.~L.~Hewett, B.~Lillie and T.~G.~Rizzo,
arXiv:hep-ph/0312193.

\end{thebibliography}
\end{document}